\documentclass[preprint,aps,showpacs,preprintnumbers,amsmath,amssymb,superscriptaddress]{revtex4-1}

\usepackage{graphicx}
\usepackage{dcolumn}
\usepackage{multirow}
\usepackage{placeins}
\usepackage{xcolor}
\usepackage[title]{appendix}
\usepackage[normalem]{ulem}

\usepackage[unicode=true, bookmarks=false, breaklinks=false, pdfborder={0 0 1},colorlinks=false]{hyperref}
\usepackage{natbib}

\newcommand{\unit}[2][1]{#1~\mathrm{#2}}

\newcommand{\KTH}{Department of Applied Physics, School of Engineering Sciences, KTH Royal Institute of Technology, AlbaNova University Center, SE-10691 Stockholm, Sweden}
\newcommand{\WISEKTH}{Wallenberg Initiative Materials Science for Sustainability (WISE), KTH Royal Institute of Technology, SE-10044 Stockholm, Sweden}
\newcommand{\SeRC}{SeRC (Swedish e-Science Research Center), KTH Royal Institute of Technology, SE-10044 Stockholm, Sweden}

\begin{document}

\title{Unveiling Mysteries of GdRu$_2$Si$_2$: 3D Magnetism in a layered like Magnet}
\author{Sagar Sarkar}
\affiliation{Department of Physics and Astronomy, Uppsala University, Uppsala, 751 20, Sweden.}

\author{Rohit Pathak}
\affiliation{Department of Physics and Astronomy, Uppsala University, Uppsala, 751 20, Sweden.}

\author{Arnob Mukherjee}
\affiliation{Department of Physics and Astronomy, Uppsala University, Uppsala, 751 20, Sweden.}

\author{Anna Delin}
    \affiliation{\KTH}
    \affiliation{\WISEKTH}
    \affiliation{\SeRC}

\author{Olle Eriksson}
\affiliation{Department of Physics and Astronomy, Uppsala University, Uppsala, 751 20, Sweden.}
\affiliation{Wallenberg Initiative Materials Science for Sustainability, Uppsala University, 75121 Uppsala, Sweden.}

\author{Vladislav Borisov}
\affiliation{Department of Physics and Astronomy, Uppsala University, Uppsala, 751 20, Sweden.}

\date{\today}

\begin{abstract}
   GdRu$_2$Si$_2$ has recently drawn significant attention as a centrosymmetric magnet capable of hosting a short-period skyrmion square lattice (SkL) in the absence of Dzyaloshinskii–Moriya interaction (DMI). In this system, Gd atoms are arranged on a square lattice, forming 2D layers separated by the Ru-Si network in the out-of-plane direction. RKKY-type exchange between the Gd moments results in an exchange frustration, which is the main source of non-collinearity in the spiral phases of the system. So far in the literature, importance has been given to in-plane $\vec{Q}$ vectors in understanding the magnetic phases of the system as they appear and have been observed on the 2D Gd layers. In this work, we calculate the Gd-Gd magnetic exchange interactions ($J_{ij}$) and perform atomic spin dynamics (ASD) simulations, providing new insights about GdRu$_2$Si$_2$. Our calculated $J_{ij}$ shows that the strongest magnetic interaction occurs between Gd atoms along the [111] body-diagonal direction of the unit cell. This, along with the body-centered tetragonal structure of the Gd sublattice, points to the presence of a hitherto ignored modulation vector, $\vec{Q}_{[111]}$, along the [111] direction in the spiral phases of the system. We confirm this with ASD simulations and establish that the magnetic phases in GdRu$_2$Si$_2$ are far more complex than they appear to be in the 2D Gd layers. This interlayer modulation vector $\vec{Q}_{[111]}$, along with the intralayer $\vec{Q}_{[100]}$ and $\vec{Q}_{[010]}$, determines the total magnetic ordering of the system. Despite its layered-like appearance both structurally and magnetically, it is in reality a strong 3D magnet, demanding additional considerations to be made for its proper theoretical modeling. Considering these, our ASD simulations provide an excellent description of the experimentally observed magnetic phase transitions. We also confirm the important role of dipolar interactions due to the large Gd moment for the correct ground-state magnetic properties. A competition between dipolar interaction and uniaxial anisotropy was found, where both try to stabilize different magnetic orderings. However, the dominance of the dipolar interaction over uniaxial anisotropy becomes evident from the ground-state properties. This indirectly points to a weak uniaxial anisotropy in this system, consistent with the spherical symmetry of the Gd $4f$ states. Our work provides a better understanding of the complex magnetism of GdRu$_2$Si$_2$. Similar important interlayer effects may also be present in many other layered magnetic systems.
\end{abstract}

\maketitle

\section{Introduction}

GdRu$_2$Si$_2$ is a magnetic metallic system that has recently attracted attention, along with some analogous systems, due to its interesting magnetic properties. The main source of magnetism here is the localized half-filled Gd $4f$ orbitals resulting in a large spin moment of $7\,\mu_B$ \cite{slaski1984magnetic}. The delocalized Gd $5d$ and Ru $4d$ states, on the other hand, create a pool of conduction electrons leading to metallicity and magnetic exchange interactions between the Gd $4f$ moments via the RKKY mechanism \cite{ruderman1954indirect}. Although Gd is a relatively heavy element, with expected large spin-orbit effects, the presence of inversion symmetry forbids the emergence of Dzyaloshinskii–Moriya interaction (DMI) [\onlinecite{DZYALOSHINSKY1958241},\onlinecite{Moriya_PhysRev_1960}], which is a widely discussed chiral interaction. As a result, the Heisenberg exchange becomes the dominant mechanism of spin-spin interactions which may potentially provide non-collinear textures in GdRu$_2$Si$_2$, as we will investigate in detail in this work. 

Like any magnetic system, the magnetic properties of GdRu$_2$Si$_2$ depend on temperature ($T$) and external magnetic field ($B$). This particular system, however, shows an interesting and rich $B-T$ phase diagram consisting of different phases \cite{khanh2020nanometric}. The primary interest is the square skyrmion lattice (SkL) phase that appears in the low-$T$ regime within a narrow range of external magnetic fields around $\unit[2]{T}$.

Magnetic skyrmions are topological solitons with a topological charge defined by the so-called skyrmion number $N_{sk}$ \cite{fert2017magnetic}. The spin ordering in a skyrmion can be considered as a special arrangement of 3D spins on a 2D plane in such a manner that they could be mapped back on the surface of a sphere \cite{fert2017magnetic,SSeki_NatCom.11_2020}. A periodic arrangement of these skyrmions is referred to as a skyrmion crystal (SkX) that shows unconventional magnetic and transport properties\cite{ANeubauer_PRL.102_2009,NKanazawa_PRL.106_2011,YShiomi_PRB.88_2013,YPMizuta_SciRep.6_2016,SSeki_NatCom.11_2020}. This makes them promising candidates for future memory devices and spintronics applications \cite{AFert_NatNTech.8_2013}. SkX is not something new and has previously been observed in chiral magnets \cite{nomoto2020formation} such as MnSi, FeGe, and Cu$_2$OSeO$_3$ \cite{MnSi1-ishikawa1976, MnSi2-muhlbauer2009, FeGe-lebech1989magnetic, FeGe2-yu2011near, Cu-1adams2012long, Cu-2seki2012observation}. In these chiral magnets, the non-collinearity of the spins in the SkX results from a competition between exchange and DMI, leading to skyrmions with a large size between 20 and 100\, nm \cite{SMuhlbauer_Science.323_2009,XZYu_Nature.465_2010}.

What makes GdRu$_2$Si$_2$ special compared to these known skyrmionic systems is that SkL results mainly from exchange frustration without DMI \cite{nomoto2020formation,bouaziz2022fermi}. This results in much smaller skyrmions of around 2 nm in the SkL phase \cite{khanh2020nanometric, bouaziz2022fermi}. The smaller size of these topological objects theoretically means a higher surface density and better device applicability, although single skyrmions not arranged in a lattice, which would be necessary for applications, have not yet been observed in GdRu$_2$Si$_2$. Hence, the ongoing efforts aim to understand, from a fundamental point of view, the magnetic properties of this system. This would aid in tuning the position of the SkL phase in the $B-T$ phase diagram. This is part of an emerging field of scientific research intending to understand the microscopic electronic and magnetic properties behind the stabilization of the SkL phase in systems without DMI. As discussed in the literature, the microscopic origin of the non-collinearity in the spiral phases is the exchange frustration due to the RKKY mechanism \cite{ruderman1954indirect}. To obtain a quantitative understanding of the exchange frustration, Nomoto  \textit{et al.} \cite{nomoto2020formation} performed an orbital decomposition of the calculated Gd-Gd exchange $J_{ij}$. 
The Fourier transform $J(q)$ of the orbital-decomposed exchange revealed an FM and AFM ground state for the $5d-5d$ and $4f-4f$ components, respectively. The presence of these competing interactions was claimed to be the reason for noncollinearity due to exchange frustration. However, in order to correctly describe the exact nature or spin configuration of the spiral states, other interactions along with exchange frustration were necessary. For example, the recent theoretical study by Boujziz \textit{et al.} \cite{bouaziz2022fermi} reported the importance of uniaxial magnetocrystalline anisotropy ($K_\mathrm{U}$) for the stability of the SkL phase in this system. Rapid progress in the GdRu$_2$Si$_2$ study made new revelations and provided new insights about the magnetic phases. Initially, the zero-field ground state of the system was reported to be a helical spiral with a single-$\textbf{Q}$ modulation\cite{khanh2020nanometric}. However, recent experimental studies with more advanced techniques confirmed a double-$\textbf{Q}$ modulated ground state\cite{NDKhanh_AdvSci.9_2022,GWood_PRB.107_2023,JSpethmann_PRM.8_2024}. For such a state, additional interactions beyond exchange frustration and uniaxial anisotropy were suggested. The importance of four-spin (biquadratic) interaction mediated by itinerant electrons in the presence of easy axis anisotropy has been claimed from the very beginning \cite{khanh2020nanometric, hayami2021topological,NDKhanh_AdvSci.9_2022} for a double-$\textbf{Q}$ phase that could transform to a topological SkL phase under a magnetic field.  It should be noted that four-spin interactions are often discussed for itinerant magnets since they are the lowest-order interactions beyond the Heisenberg exchange that can be significant enough to affect the magnetic order. Though a consensus has not yet been achieved as to what drives the skyrmion formation, there is a general agreement regarding the characterization of different magnetic phases in terms of the modulation vectors. The two in-plane modulation vectors $\textbf{Q}_{100}$ along $\textbf{a}$-axis and $\textbf{Q}_{010}$ along $\textbf{b}$-axis have been considered important and are enough to describe the magnetic phases. This is because each phase appears and has been observed in the two-dimensional (2D) layers of the Gd square lattice.

In this report, we calculate and analyze in detail the magnetic exchange interactions in this layered rare-earth system GdRu$_2$Si$_2$. We show that the magnetic phases are more complex than what was discussed in the literature so far and cannot be described with the in-plane modulation vectors alone. The calculated exchange interactions indicate, for example, that the exchange frustration results from competition between the intralayer and interlayer exchange, where the latter was mostly neglected in previous studies. A stronger interlayer FM exchange along with the body-centered tetragonal Gd sublattice results in an interlayer modulation vector $\textbf{Q}_{111}$ along the [111] body-diagonal direction of the unit cell which we confirm using atomistic spin dynamics (ASD) simulations. The $\textbf{Q}_{111}$ modulation of the spin texture cannot be neglected while simulating the magnetic phase diagram, since ASD simulation for a supercell incompatible with $\textbf{Q}_{111}$ does not show the experimentally observed magnetic phase transitions as a function of applied magnetic field $\textbf{B}$. From these simulations, we confirm the importance of dipolar interactions that govern the ground-state magnetic properties of the system.  A competing effect of the dipolar interaction and uniaxial anisotropy was found, where both try to stabilize different magnetic orderings. The dominance of the dipolar interaction over uniaxial anisotropy becomes evident from the ground-state properties. This indirectly points to a weak uniaxial anisotropy consistent with the spherical symmetry of the Gd $4f$ states. However, stabilization of the critical SkL phase might still require other weak interactions like biquadratic and symmetric anisotropic exchange acting as a perturbation.


While our work is about a concrete magnetic system, the discussion of magnetic phases and theoretical approaches for their simulation is quite general and can be applied to many other noncollinear magnets with spiral or skyrmion phases. In particular, using GdRu$_2$Si$_2$ as an example, we point out some important aspects of our methodology that are required to correctly model the magnetic phase diagram using atomistic spin dynamics (ASD). Since ASD is widely used nowadays for studying magnets, the methodological conclusions of our study can make a useful contribution to the field in general. It is also worthwhile to mention that GdRu$_2$Si$_2$ has a very common type of crystal structure, the so-called 122- or ThCr$_2$Si$_2$-type of structure, which is also found in widely discussed iron-based superconductors and magnets \cite{CaFe2As2} and rare-earth magnets \textit{Ln}Rh$_2$Si$_2$ with ultrafast demagnetization dynamics \cite{Windsor2022} (see further examples in \cite{Hoffmann1985}).

\section{Methodology}

\subsection{Electronic structure and Magnetic exchange}

We have considered the two-formula unit tetragonal unit cell for our system. The lattice parameters were kept fixed at the experimental values \cite{KHiebl_JMMM.37_1983}. First, we have calculated the electronic structure from density functional theory (DFT) using a projected augmented wave (PAW) method~\cite{blochl_PhysRevB.50.17953_1994,kresse_PhysRevB.59.1758_1999} as implemented in the Vienna Ab initio Simulation Package 
(VASP)~\cite{kresse_PhysRevB.47.558_1993,kresse_PhysRevB.49.14251_1994,kresse_PhysRevB.54.11169_1996,kresse_cms.6.15_1996}. The generalized gradient approximation (GGA)~\cite{perdew_PhysRevLett.77.3865_1996} in the Perdew-Burke-Ernzerhof (PBE) parametrization~\cite{perdew_PhysRevLett.77.3865_1996,perdew_PhysRevLett.78.1396_1997} was considered for the exchange-correlation functional and a plane-wave energy cutoff of $\unit[500]{eV}$ was used for the basis set. Along with this, a $\Gamma$-centered
Monkhorst-Pack 20$\times$20$\times$10 {\bf{k}}-mesh, providing convergence of the total energy and local moments, was used for reciprocal space integration. The main purpose was to check the effect of the crystalline environment and other delocalized $s$, $p$, and $d$ states on the atomic-like localized Gd $4f$ states. For this, the Gd $4f$ states were considered as part of the valence electronic manifold, and to achieve a correct description of their localized nature, we have performed a DFT+$U$ calculation within the Hartree-Fock approximation~\cite{anisimov_jpcm.9.48_1997,kotliar_RevModPhys.78.865_2006}. For this, the rotationally invariant formulation of Liechtenstein {\it{et al.}}~\cite{liechtenstein_PhysRevB.52.R5467_1995} was used with the Coulomb interaction parameters $U = \unit[6.7]{eV}$ and $J = \unit[0.7]{eV}$ on the Gd $4f$ states, based on a previous study~\cite{TNomoto_JAP.133_2023}. Calculations based on this methodology were used here for an initial analysis of electronic properties presented also in Fig.~\ref{electronic}.

Following this, to extract the magnetic exchange interactions between the Gd moments,  the electronic structure was again calculated with the Full Potential Linear Muffin-Tin Orbital (FP-LMTO) method, as implemented in the RSPt
software\cite{Wills1987,Wills2000,wills2010full,RSPt}. GGA-PBE exchange-correlation functional was used similarly to the VASP calculations. These calculations included two sets of basis functions, covering both valence and semi-core states. These states were specifically constructed from the $6s$, $6p$, and $5d$ orbitals for Gd and Ru, and the $3s$, $3p$, and $3d$ orbitals for Se, respectively. The Gd $4f$ states were considered in the core to be treated scalar relativistically to reduce the computational cost. Hence, a DFT+$U$ calculation was not required in this case. We chose kinetic tail energies as $-0.1$, $-2.3$, and $1.5$ Ry.
For the Brillouin zone sampling, we used an optimized $\gamma$-centered Monkhorst-Pack mesh of 32$\times$32$\times$16 {\bf{k}}-points. From here, the \textit{ab-initio} Kohn-Sham Hamiltonian or the DFT Hamiltonian is mapped onto an effective Heisenberg Hamiltonian with classical spins of the following form \cite{LIECHTENSTEIN198765,Igor_PhysRevB_2015} to extract the interatomic exchange interactions through the application of the magnetic force theorem (MFT) \cite{LIECHTENSTEIN198765,Lichtenstein_PhysRevB_2000} (for a review, see Ref. \cite{Szilva2023}):
\begin{equation}
 {H} = - \sum_{i\neq j} J_{ij}\, \vec{e}_{i}\cdot\vec{e}_{j}.
 \label{eqn1}
\end{equation}
Here, $(i,j)$ are the indices for the magnetic sites in the system, while $\vec{e}_i$ and $\vec{e}_j$ are the unit vectors along the spin directions at the sites $i$, and $j$, respectively; $J_{ij}$ is the exchange interaction between the two spins at the sites $i$ and $j$. In the Green-function-based approach to evaluate the $J_{ij}$ values used in the RSPt software, the exchange parameters $J_{ij}$ are determined from a generalized non-relativistic expression as given below~\cite{Igor_PhysRevB_2015}.

\begin{equation}
J_{ij}=\frac{T}{4} \sum_{n} \operatorname{Tr}\left[\hat{\Delta}_{i}\left(i \omega_{n}\right)\hat{G}_{ij}^{\uparrow}\left(i \omega_{n}\right) \hat{\Delta}_{j}\left(i \omega_{n}\right) \hat{G}_{ji}^{\downarrow}\left(i \omega_{n}\right)\right].
 \label{eqn2}
\end{equation}
Here $T$ is the temperature, and $\hat\Delta$ is the onsite exchange potential giving the exchange splitting at sites $i$ and $j$; $\hat{G}_{ij}^{\sigma}$ is the intersite Green's function projected over spin $\sigma$ that can have values $\{\uparrow , \downarrow\}$ and $\omega_n$ is the $n^\mathrm{th}$ fermionic Matsubara frequency. All the terms in the above expression are matrices in orbital and spin space with the trace running over the orbital degrees of freedom. The summation is over Matsubara frequencies ($\omega_n$). To ensure precise convergence in our analysis of exchange parameters, we increased the $k$-mesh to $52\times 52\times 26$ for these calculations. We note that the exchange interactions in this system are very similar within the non-relativistic and fully-relativistic limits (see a comparison in section~II of the SM).

We employed the force theorem to determine the magnetic anisotropy, as this method has greater efficiency and accuracy from a computational point of view compared to the method based on total magnetic energies for different orientations of the global spin axis, as discussed, e.g., in previous works \cite{MAE-1, MAE-2}.
In this approach, we first conducted non-relativistic self-consistent DFT calculations using the RSPt code. Once we obtained the converged potential and charge density, we performed three fully relativistic non-self-consistent (single-iteration) calculations with magnetic moments aligned along the three mutually orthogonal [100], [010], and [001] Cartesian directions. The magnetic anisotropy energy (MAE) was then calculated by taking the difference between the eigenvalue sums for the in-plane and out-of-plane magnetic moment directions. For the MAE calculations, we used a converged $60\times 60\times 30$ $k$-mesh, slightly finer than for the $J_{ij}$ calculation.

\subsection{Atomistic Spin Dynamics Simulations}

In the next step, spin textures were simulated by atomistic spin dynamics (ASD) using the Uppsala Atomistic Spin Dynamics (UppASD) package [\onlinecite{uppasd},\onlinecite{Eriksson2017}], where also the inclusion of spin-lattice coupling effects can be done in a systematic first-principles way following the methodology in \cite{Hellsvik2019}. 
In this work, we focus on the spin part of the system and solve the Landau-Lifshitz-Gilbert (LLG) equation [\onlinecite{Landau1935},\onlinecite{Gilbert2004}] for the atomic magnetic moments:
\begin{equation}
    \frac{d\Vec{m_i}}{dt} = - \frac{\gamma}{1+\alpha^2}\, \Vec{m}_i \times [\Vec{B}_i + \Vec{b}_i(t)] - \frac{\gamma}{m_i} \frac{\alpha}{1+\alpha^2}\, \Vec{m}_i \times (\Vec{m}_i \times [\Vec{B}_i + \Vec{b}_i(t)]).
     \label{eqn3}
\end{equation}
Here, $\gamma$ is the gyromagnetic ratio, and $\Vec{b}_i(t)$ is a stochastic magnetic field with Gaussian distribution. The magnitude of this field is related to the damping parameter $\alpha$, which helps bring the system into thermal equilibrium, and temperature $T$. We use a time step of $\Delta t = \unit[0.1]{fs}$ for the annealing phase and $\Delta t = \unit[1]{fs}$ for the measurement phase in the UppASD calculations to solve these differential equations.

The effective field $\Vec{B}_i$ experienced by each spin $i$ is derived from the partial derivative of the Hamiltonian $H$ with respect to the local magnetic moment,
\begin{equation}
    \Vec{B}_i = - \frac{\partial H}{\partial \Vec{m}_i}.
     \label{eqn4}
\end{equation}
The Hamiltonian $H$ includes all relevant interactions and is given by:
\begin{equation}
    H = - \frac{1}{2} \sum_{i \neq j} J_{ij}\, \vec{e}_i \cdot \vec{e}_j - K_U \sum_{i} \left( \vec{e}_i \cdot \vec{z}\, \right)^2 - \sum_{i} \vec{B}_\mathrm{ext} \cdot \vec{e}_i,
     \label{eqn5}
\end{equation}
where the first term describes the Heisenberg exchange, the second term is the uniaxial anisotropy, and the final term corresponds to the effect of an external field $\vec{B}_\mathrm{ext}$. For small $\vec{B}_\mathrm{ext}$, the most significant contribution to the Hamiltonian is typically from Heisenberg exchange interaction, where $i$ and $j$ are atomic indices, and $J_{ij}$ is the strength of the exchange interaction, obtained from our first-principles calculations. For a magnetic system like GdRu$_2$Si$_2$ with large atomic spins of $7 \mu_B$, dipolar interactions could become important. For this, the following interaction term $H_{dip}$ was added to our Hamiltonian in Eq.~\ref{eqn5} as required. 

\begin{equation}
    H_{dip} = - \frac{\mu_0}{4\pi} \sum_{i,j,j \neq i} \frac{1}{r^3_{ij}}[3(\textbf{m}_i \cdot \hat{r}_{ij}) (\textbf{m}_j \cdot \hat{r}_{ij}) - \textbf{m}_i \cdot \textbf{m}_j],
     \label{eqn6}
\end{equation}
where $\textbf{m}_i$, $\textbf{m}_j$ are the atomic moments at sites $i$ and $j$, respectively, and $\mu_0$ is the magnetic constant.

In the atomistic spin dynamics (ASD) simulation, we used a periodic simulation box with a size of, for example, $37 \times 37 \times 37$ conventional unit cells with 2 Gd atoms per cell, resulting in 101306 spins. We first performed a simulated annealing to bring the system into thermal equilibrium, followed by an ASD measurement phase to obtain the spin texture after the system had evolved via the LLG equation and reached an energy minimum. Simulated annealing was performed at gradually decreasing temperatures of $\unit[200]{K}$, $\unit[100]{K}$, $\unit[50]{K}$, and $\unit[10]{K}$, with 20,000 spin-dynamics sampling steps at each temperature. After these annealing steps, we performed 500,000 sampling steps during the measurement phase, so that the spin system can reach an equilibrium state at zero temperature and a given external magnetic field.

\section{Results and discussion}

\subsection{Structural Properties}

\begin{figure}[ht]
\includegraphics[scale=0.5]{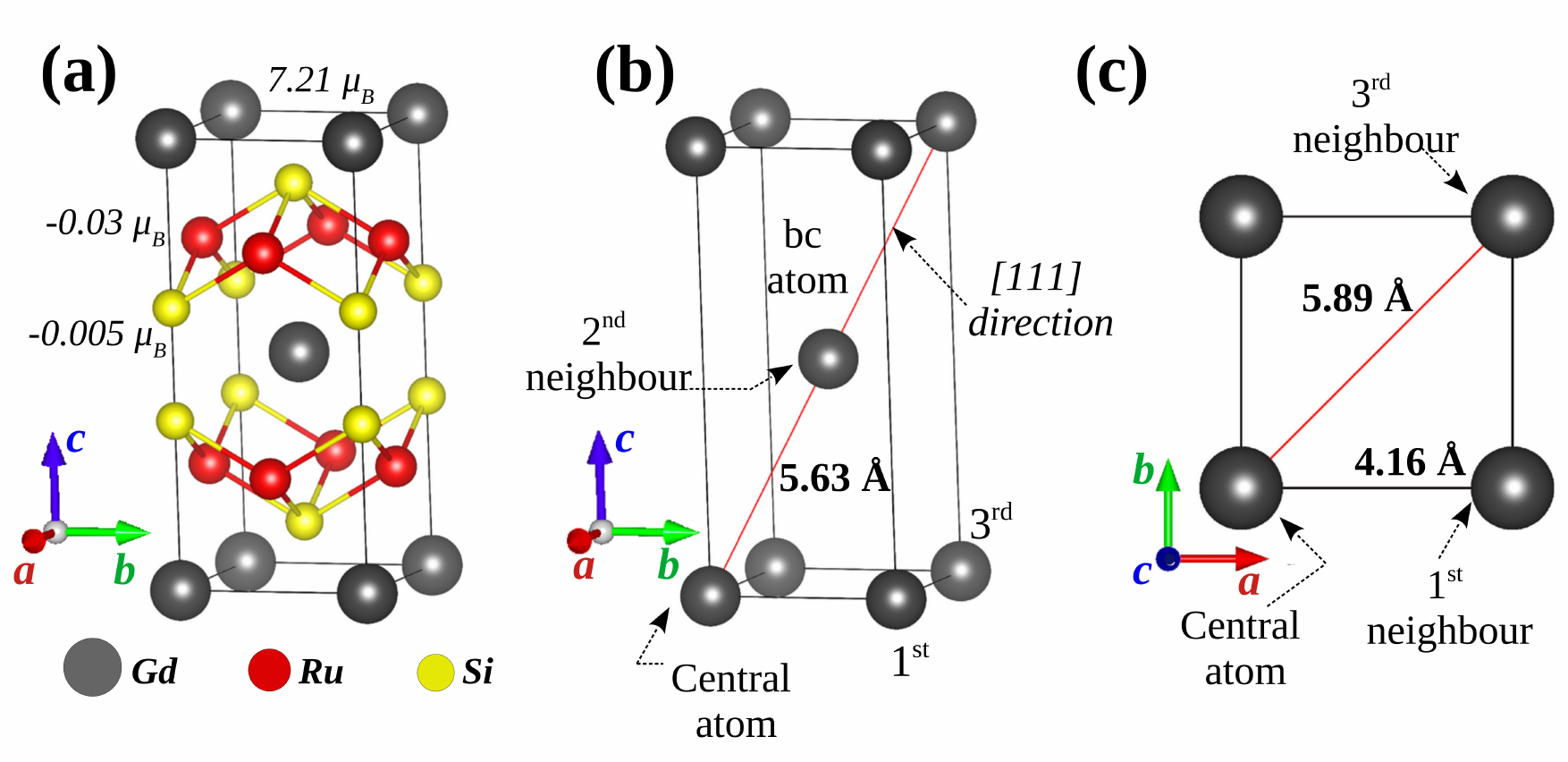}
\caption{(a) The experimental tetragonal unit cell of GdRu$_2$Si$_2$ with almost a layered structure. The 2D Gd layers are separated from each other by the Ru-Si networks along the $c$-axis. The magnetic moments of each atom are also indicated, which were obtained from DFT calculations described in section B. (b) Position of the Gd atoms in the unit cell showing the formation of a body-centered tetragonal sublattice. The second neighbor direction and the corresponding Gd-Gd distance are shown. (c) Square lattice arrangement of the Gd atoms in the $ab$-plane of a single Gd layer. The in-plane nearest neighbor and third neighbor directions and corresponding Gd-Gd distances are shown.}
\label{structure}
\end{figure}

We start with the structural details of GdRu$_2$Si$_2$, and the experimental unit cell~\cite{KHiebl_JMMM.37_1983} that is shown in Fig.~\ref{structure} (a). In this system, the Gd atoms are the main source of magnetism. The Gd sublattice forms a body-centered tetragonal structure (bct), as shown in Fig.~\ref{structure} (b). Hence, we can see that the structure becomes quite interesting with layers of Gd atoms stacked along the $\textbf{c}$-axis and separated from each other by the metallic Ru-Si networks. This initially gives an impression of a layered quasi-two-dimensional magnetic structure. However, that is far from reality, as we show in the present work. This will be evident when we discuss the magnetic properties of the system. Each Gd layer separated by the Ru-Si network discussed above forms a 2D square lattice, as shown in Fig.~\ref{structure} (c). The tetragonality ratio $c/a$ of the system is such that the Gd atom in the body-centered position along the [111] direction of the unit cell becomes the second neighbor atom for the central atom, as indicated in Fig.~\ref{structure} (b). The third neighbor atom lies along the [110] in-plane direction of the square lattice, as pointed out in Fig.~\ref{structure} (c). The first, second, and third neighbors are shown here in the structural description as they become important in understanding the magnetic exchange and the magnetic properties. 

\FloatBarrier

\subsection{Electronic Properties}

\begin{figure}[ht]
\includegraphics[scale=0.6]{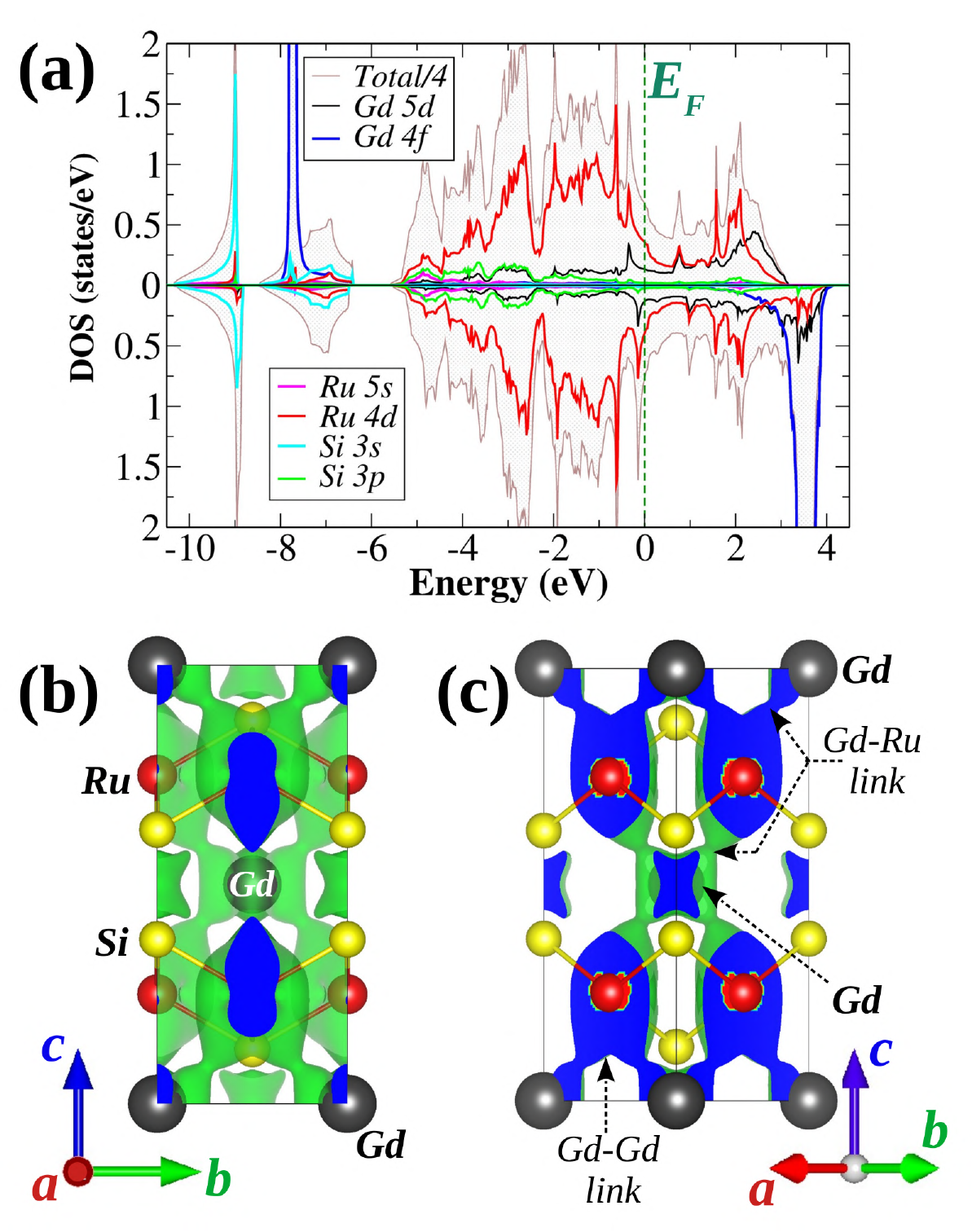}
\caption{(a) Total (scaled down by 4) and atom+orbital projected density of states from VASP calculation where Gd $4f$ states were treated as valence states. See the main text for details. (b) and (c) show the charge density distribution corresponding to the valence states around Fermi energy along [100] and [110] directions, respectively. See the main text for more details.}
\label{electronic}
\end{figure}

Next, we briefly discuss the basic electronic properties of the system. For this, we consider the unit cell shown in Fig.~\ref{structure} (a) with FM ordering of the Gd spins. Non-relativistic GGA+$U$ calculation was performed in VASP considering the Gd $4f$ states in the valence with the values of $U$ and $J$ equal to 6.70, and 0.70~eV respectively ~\cite{TNomoto_JAP.133_2023}.  The aim was to check the interaction between the $4f$ and other valence states of the system. The calculated total density of states (DOS) and atom-projected partial density of states (PDOS) are shown in Fig.~\ref{electronic} (a). From these plots, we can see that the system is metallic. Although we have an FM order, the exchange splitting of the states near the Fermi energy ($E_\mathrm{F}$) is weak. This results in tiny induced moments on the Ru and Si atoms, as indicated in Fig.~\ref{structure} (a). Large exchange splitting due to FM order is visible only for the localized Gd $4f$ states (blue lines) far away from $E_\mathrm{F}$ and resulting in a large spin of $7 \mu_B$. Major contributions at $E_\mathrm{F}$ come from the Ru $4d$ (about 65\%) and Gd $5d$ (about 30\%) states, as shown by the red and black lines, respectively. Si $3p$ states (green line) have a minimal contribution around $E_\mathrm{F}$ and can be ignored. These states form a pool of conduction electrons distributed within the system. To verify the nature of this distribution, a small energy window of 0.20~eV was considered around the Fermi level, and the charge density (CD) corresponding to these states was calculated as shown in Figs.~\ref{electronic} (b) and (c) from two different angles. An isosurface level was selected in such a way as to make the CD visible around both the Ru and Gd atoms. The charge density does not show a layered-type structure localized in the individual Gd and Ru-Si atomic layers. We can see prominent links connecting the charge densities around the Gd and Ru atoms, as shown in Fig.~\ref{electronic} (c).
Most importantly, a strong intralayer contact between the Gd atoms via the CD is missing. Instead, the Gd atoms get connected indirectly via the Ru atoms. On the other hand, the conduction electrons are the main source in this system for mediating magnetic exchange between Gd atoms, and these interactions have been suggested to have an RKKY character~\cite{ruderman1954indirect}.  For that reason, the anisotropic structure of the conduction electron density is expected to impact the magnetic exchange interactions which we discuss in the next section.

\FloatBarrier

\subsection{Magnetic Properties}

\subsubsection{Magnetic exchange and 3D magnetism of GdRu$_2$Si$_2$}

\begin{figure}[ht]
\includegraphics[scale=0.55]{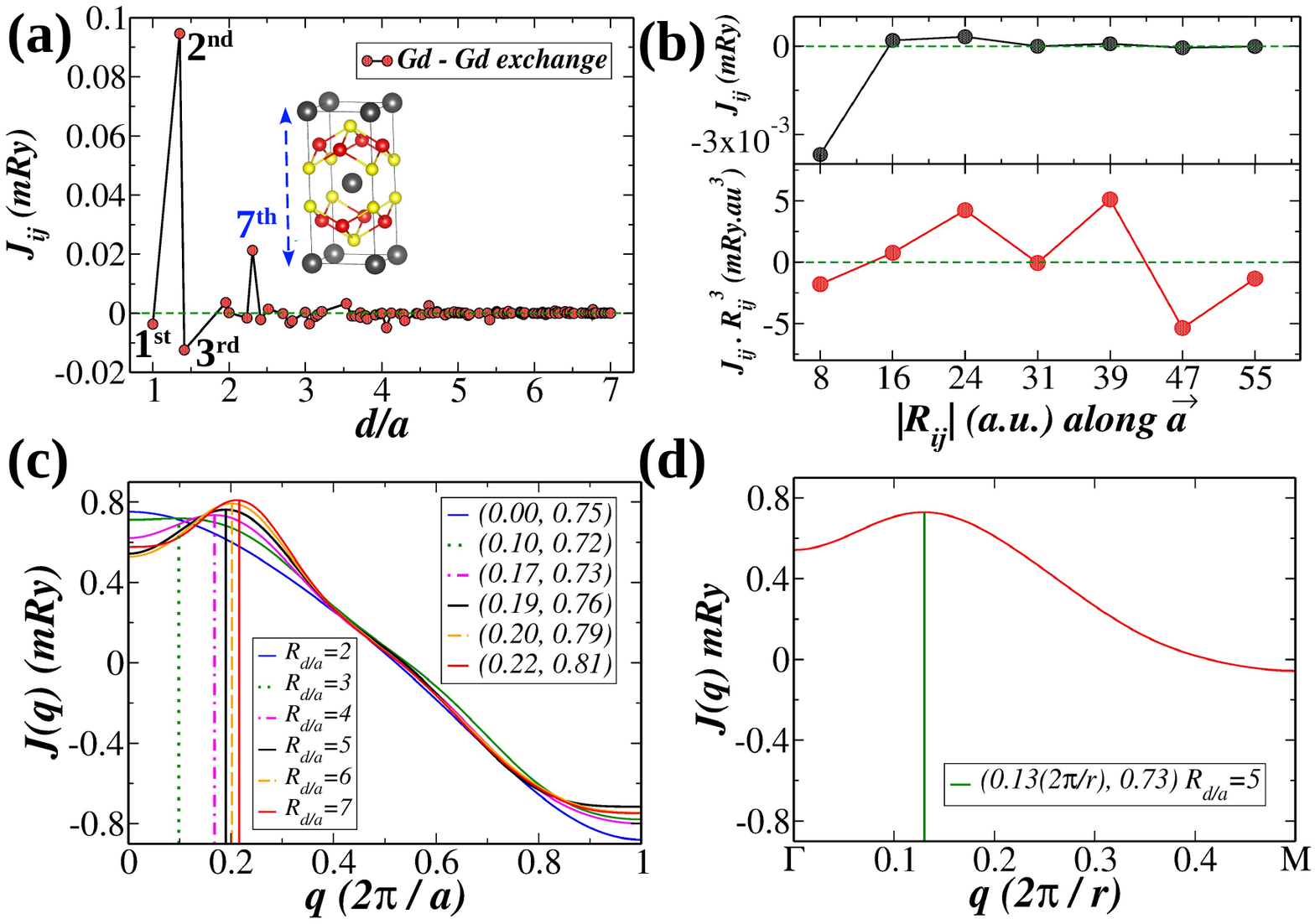}
\caption{(a) Calculated Gd-Gd interatomic exchange interactions as functions of distance ($d$) scaled by the lattice constant ($a$). The inset depicts the exchange path corresponding to the $7^\mathrm{th}$ interaction in the out-of-plane direction. (b) As functions of distance $R_{ij}$ along the crystallographic $a$-axis, $J_{ij}$ and $J_{ij}\cdot R^{3}_{ij}$ reveal oscillations which point at the RKKY exchange mechanism. (c) Fourier transforms $J(q)$ of the calculated exchange $J_{ij}$ for $\textbf{q}$ along $\textbf{a}$ in the unit of $2\pi/a$ for different values of the cutoff radii $R_{d/a}$. The inset shows the ($|\textbf{Q}_{100}|$, $J(|\textbf{Q}_{100}|$)) values corresponding to the maxima of $J(q)$ (see the main text for details). (d)  Fourier transform $J(q)$ of the calculated exchange $J_{ij}$ for $\textbf{q}$ along $\Gamma-M$ direction of the Brillouin Zone with $R_{d/a} = 5$. $\Gamma-M$ direction is $||$ to the [110] crystallographic direction. 
} 
\label{magnetic}
\end{figure}

In the next step, we calculated the magnetic exchange interactions for the FM reference state via magnetic force theorem (MFT) \cite{LIECHTENSTEIN198765,Lichtenstein_PhysRevB_2000} (review in \cite{Szilva2023}) available in the RSPt software [\onlinecite{Wills1987,Wills2000,wills2010full}]. The Gd $4f$ states were treated as spin-polarized core states in this calculation since they represent non-hybridized, localized electron states, a configuration of the $4f$ states that closely represents the so-called Standard Model of the rare-earths \cite{RareEarthBook}. Note that a more precise treatment would involve the so-called Hubbard-I approximation, which was investigated for the rare-earth elements in detail in Ref.~\cite{Locht2016}. However, for Gd-based systems, where the spin-up $4f$ shell is filled and spherically symmetric, treating the $4f$ states as a spin-polarized core or a spin-polarized valence band closely resembles the Hubbard-I approximation, which motivates the approximations used here. A comparison between VASP and RSPt calculated valence electronic states around the Fermi energy $E_\mathrm{F}$ is shown in Fig.~S1 from section~I of the SM. Magnetic interactions ($J_{ij}$) between Gd moments at different distances ($d$) (scaled by the lattice constant $a$) are shown in Fig.~\ref{magnetic} (a). We see a nearly alternating FM and AFM coupling with the neighboring atoms reminiscent of a free carrier-mediated RKKY type exchange~\cite{ruderman1954indirect}, a well-known mechanism in intrinsic\cite{YKvashnin_PRL.116_2016} and extrinsic\cite{SSarkar_2Dmat.9_2022,SSarkar_PRB.110_2024} metallic magnets. RKKY has also been claimed in recent studies as the primary mechanism for magnetic interactions in this system~\cite{nomoto2020formation, YYasui_Natcom.11_2020,bouaziz2022fermi}. To confirm this, we have separately plotted $J_{ij}$ as a function of the interatomic distance $R_{ij}$ along the in-plane $\textbf{a}$-axis, as shown in the first panel of Fig.~\ref{magnetic} (b). Note that $J_{ij}$($R_{ij}$) decays quickly, making it impossible to view its oscillatory properties. To overcome this and take into account the estimated scaling factor of $1/R^3_{ij}$ for RKKY interaction in metallic 3D systems\cite{NDAristov_PRB.55_1997},  we plot $J_{ij} \cdot R^3_{ij}$ as a function of $R_{ij}$ as shown in the bottom panel of Fig.~\ref{magnetic} (b). The resulting quantity $J_{ij} \cdot R^3_{ij}$ shows prominent oscillatory behavior revealing the RKKY-type exchange mechanism present in the system. Hence, long-ranged interatomic exchange with non-negligible interaction strength is not unexpected in this system as can be seen from Fig.~\ref{magnetic} (a), where the 2$^\mathrm{nd}$ ($J_2$) and 7$^\mathrm{th}$ ($J_7$) interactions come out to be stronger than the 1$^\mathrm{st}$ ($J_1$) and 3$^\mathrm{rd}$ ($J_3$) interactions. These four interactions are important and are discussed below in more detail.  The 1$^\mathrm{st}$, 2$^\mathrm{nd}$, and 3$^\mathrm{rd}$ neighbor exchange paths are shown in Fig.~\ref{structure} and have been discussed in the previous sections. The 7$^\mathrm{th}$ neighbor exchange path is shown in the inset of Fig.~\ref{magnetic} (a), which is along the $\textbf{c}$-axis in the out-of-plane directions. To summarize, $J_1$ and $J_3$ are two intralayer exchange interactions that are weakly AFM, while $J_2$ and $J_7$, on the other hand, are two interlayer interactions that are FM in nature and stronger than the intralayer interactions. A stronger interlayer exchange suggests that the system should not be treated as a layered 2D magnetic system. Interlayer exchange, if considered alone, will stabilize an FM ordering without any magnetic frustration. However, the two intralayer AFM exchange interactions on the square lattice by symmetry may lead to an exchange frustration, and an overall non-collinear state could be found. This is in line with the experimental observation of non-collinear spiral states within the 2D Gd layers\cite{khanh2020nanometric,NDKhanh_AdvSci.9_2022,GWood_PRB.107_2023,JSpethmann_PRM.8_2024}.  These states could be well described\cite{hayami2021square,NDKhanh_AdvSci.9_2022} by the two in-plane (IP) spin spiral modulation vectors $\textbf{Q}_{100}$ and $\textbf{Q}_{010}$ with the experimentally reported value  $\textbf{Q}^{\text{Exp}}_{100}$ = (0.22, 0, 0) along the $\textbf{a}$-axis in the unit of reciprocal lattice vector $\textbf{a}^*$~\cite{khanh2020nanometric}. As the zero field spiral GS should mainly result from the Heisenberg exchange frustration in the absence of any DMI, a Fourier transform $J(q)$ of our calculated $J_{ij}$'s or the adiabatic magnon spectra (AMS) can predict such a modulation vector. This analysis, in turn, may confirm or refute the correctness of our calculated exchange interactions.  In the $J(q)$ calculation, $R_{d/a}$ is a parameter defining the cut-off radii of a sphere around any magnetic site, measured in the unit of the lattice constant $a$. Only $J_{ij}$'s between that magnetic site and its neighbors inside the sphere are considered for the calculation of $J(q)$, and hence, it is expected to change with $R_{d/a}$. In Fig.~\ref{magnetic} (c), we show $J(q)$ for $\textbf{q} \parallel \textbf{a}$ in the units of reciprocal lattice vector $\textbf{a}^*$, for different $R_{d/a}$ values. One can see that an FM stability only occurs for $R_{d/a}$ = 2 and becomes unstable for $R_{d/a} \ge 3$ where a spin-spiral state with a modulation vector $\textbf{Q}_{100}$ becomes favorable. $|\textbf{Q}_{100}|$ is identified by the value of $q$ where $J(q)$ reaches a maximum. For each considered $R_{d/a}$, the values ($|\textbf{Q}_{100}|$, $J(|\textbf{Q}_{100}|$)) are shown in the inset of Fig.~\ref{magnetic} (c). $\textbf{Q}_{100}$ increases in magnitude gradually and becomes equal to the experimentally reported $\textbf{Q}^{\text{Exp}}_{100}$ for $R_{d/a} = 7$. But since we can see from Fig.~\ref{magnetic} (a) that most of the calculated $J_{ij}$'s for $d/a \ge 5$ are nearly equal to zero, we have considered $R_{d/a} = 5$ for which  $J(q)$ also predicts a spin-spiral modulation vector (0.19, 0, 0) very close to $\textbf{Q}^{\text{Exp}}_{100}$. For a better understanding of the possible exchange spirals, we have also examined  $J(q)$ for $\textbf{q}$ $\parallel$ different high-symmetry (HS) directions of the Brillouin Zone (see Fig.~S3 of SM to view the BZ) and is presented in Fig.~S4 of the SM. We found that $J(q)$ for $\textbf{q}$ along $\Gamma-Y \parallel \textbf{b}$ is identical to $\textbf{q} \parallel \textbf{a}$ showing $|\textbf{Q}_{010}| = |\textbf{Q}_{100}|$ with the same stabilization energy of 0.76 mRy/spin. This happens due to the in-plane square lattice symmetry of the Gd layers. These two vectors $\textbf{Q}_{100}$ and $\textbf{Q}_{010}$ also show the maximum stability over all other vectors.  A close competition only comes from $\textbf{Q}_{110}$ along $\Gamma-M$ $\parallel$ to the in-plane crystallographic direction [110], for which the stabilization energy is 0.73 mRy/spin shown in Fig.~\ref{magnetic} (d). In summary, this suggests that the resulting zero-field exchange spiral should be governed by $\textbf{Q}_{100}$ or $\textbf{Q}_{010}$ or both, which is consistent with all experimental observations\cite{khanh2020nanometric,NDKhanh_AdvSci.9_2022,GWood_PRB.107_2023,JSpethmann_PRM.8_2024} made so far.

\FloatBarrier

\begin{figure}[ht]
\includegraphics[scale=0.5]{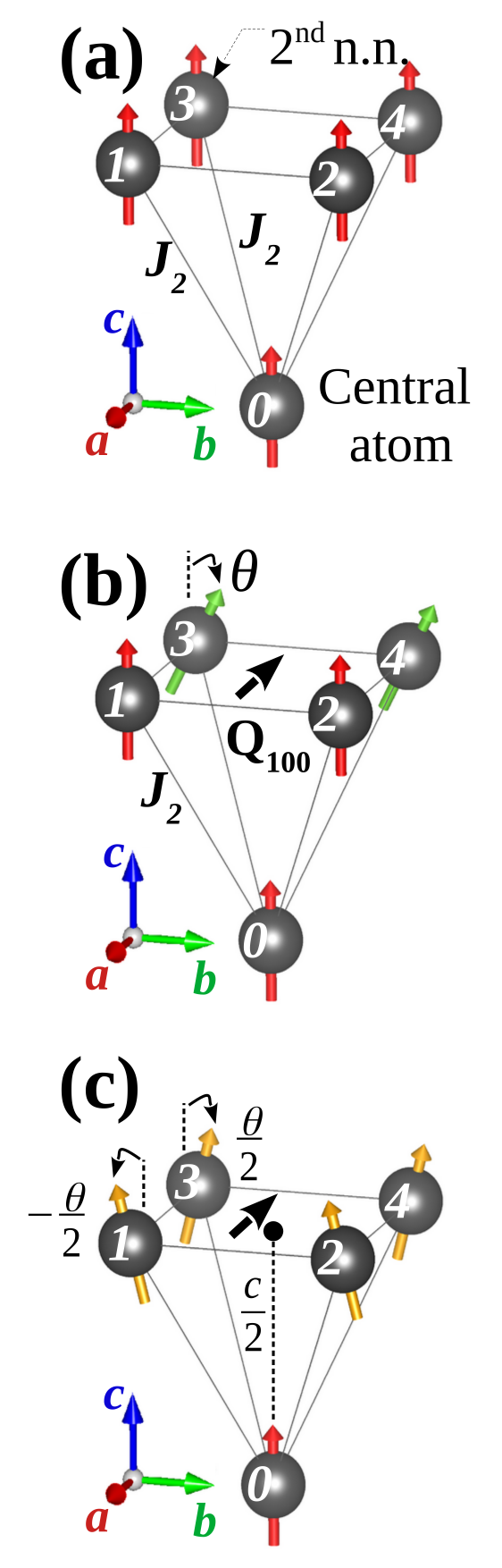}
\caption{(a) The central Gd atom from the unit cell origin and its four second nearest neighbors (n.n.) along the [111] directions on the top Gd layer with an FM coupling due to the exchange $J_2$. (b) A phase difference $\theta$ between the four Gd spins on the top layer along the $a$-axis in a helical phase due to the presence of the modulation vector $\textbf{Q}_{100}$. (c) The same helical phase but with a slightly different spin configuration on the top layer.
}
\label{helical}
\end{figure}

The two in-plane vectors  $\textbf{Q}_{100}$ and $\textbf{Q}_{010}$ have been considered important entities to characterize and understand the magnetic phases of GdRu$_2$Si$_2$. The obvious reason is that the spin configurations corresponding to different magnetic phases appear and have been observed within the 2D plane of the Gd square lattice. This, together with the layered type structure of GdRu$_2$Si$_2$ creates an impression of a 2D magnetic system. The interlayer coupling between the Gd layers was also considered weakly FM ~\cite{bouaziz2022fermi} having no major role in the magnetic properties. Our calculations, however, suggest that this is not true as we find the FM interlayer exchange to be the strongest magnetic interaction in the system, even stronger than the nearest-neighbor intralayer AFM couplings, which means that GdRu$_2$Si$_2$ is essentially a 3D magnet. This, together with the \textit{bct} sublattice of the Gd atoms, makes the magnetic phases much more complex than they appear to be with inter-layer magnetic coupling and spin-spin correlations. This interlayer correlation could be understood from the following discussion on the role of the strongest FM exchange between layers $J_{2}$ in any possible spin spiral state of the system. In this regard, we recall that the zero-field ground state was initially reported to be a helical spiral\cite{khanh2020nanometric} with a single modulation vector $\textbf{Q}_{100}$, but more recent experiments have revealed a double-$\textbf{Q}$ modulated state\cite{NDKhanh_AdvSci.9_2022,GWood_PRB.107_2023,JSpethmann_PRM.8_2024} with both $\textbf{Q}_{100}$ and $\textbf{Q}_{010}$. For simplicity of analysis, in order to show the importance of the interlayer $J_{2}$ exchange coupling, we consider a single-$\textbf{Q}$ helical state as a reference state here. However, this analysis and the conclusions drawn are valid for any single or double-$\textbf{Q}$ reference state.  Fig.~\ref{helical} (a) shows the central Gd atom (0$^\mathrm{th}$) from the unit cell origin and its four 2$^\mathrm{nd}$ nearest neighbors (1 to 4) in the next Gd layer forming the square lattice.  An FM ordering between the 0$^\mathrm{th}$ atom and its four 2$^\mathrm{nd}$ nearest neighbors is shown considering the ferromagnetic $J_{2}$ interaction only. This is true only in the absence of intralayer exchange, which should otherwise induce a spiral ground state. Assuming this state to be a helical state with a single modulation vector $\textbf{Q}_{100}$, the four spins on the top layer in Fig.~\ref{helical} (a) cannot remain in phase as shown in Fig.~\ref{helical} (b). A phase difference of $\theta = \bf{Q_{100}\cdot a}$ in the $3^\mathrm{rd}$ and $4^\mathrm{th}$ spins with respect to the $1^\mathrm{st}$ and $2^\mathrm{nd}$ spins results from the $\textbf{Q}_{100}$-modulation. Due to this, it is evident that the $0^\mathrm{th}$ spin cannot remain in phase with the four spins on the top layer simultaneously. From Fig.~\ref{helical} (b) we can see that if it stays in phase with the $1^\mathrm{st}$ and $2^\mathrm{nd}$ spins, a phase difference of $\theta$ must occur with the $3^\mathrm{rd}$ and $4^\mathrm{th}$ spins. The helical state shown in Fig.~\ref{helical} (b) can also be realized with a slightly different spin configuration as depicted in Fig.~\ref{helical} (c). If we ignore the 0$^\mathrm{th}$ spin and $J_{2}$ exchange coupling, then both Fig.~\ref{helical} (b) and Fig.~\ref{helical} (c) represent the same magnetic state (helical state) on the top layer and are energetically degenerate. In both cases, the phase difference between two neighboring spins along $\textbf{a}$ is $\theta$ due to $\textbf{Q}_{100}$. However, if we consider the interlayer interaction $J_2$ with the 0$^\mathrm{th}$ spin, then the configuration in Fig.~\ref{helical} (c) shall become energetically more favorable over the configuration in Fig.~\ref{helical} (b). This can be seen from the energy differences between the spin configurations in Fig.~\ref{helical}a-c. If we consider unit spins on the Gd atoms, then the energy difference between configurations (b) and (a) is $E_b - E_a = 4J_{2} (1-\mathrm{cos}^{2}(\theta/2)) $, whereas for the configurations (c) and (a) $E_c - E_a = 4J_{2} (1-\mathrm{cos}(\theta/2)) $. Now, for any small value of the spin canting angle $\theta$, $\mathrm{cos}^{2}(\theta/2) < \mathrm{cos}(\theta/2)$, and hence $E_c < E_b$. So the spiral spin configuration in Fig.~\ref{helical} (c) is energetically preferable.

\FloatBarrier

 This analysis shows an important spin correlation between the second neighbor spins along the body diagonal direction ($\textbf{a} + \textbf{b} + \textbf{c}$) or [111]. If we have an intralayer phase difference of $\theta$ between two neighbors along $\textbf{a}$ or $\textbf{b}$ due to $\textbf{Q}_{100}$ or $\textbf{Q}_{010}$, then an interlayer phase difference of $\theta/2$ must occur between two neighboring spins along ($\textbf{a} + \textbf{b} + \textbf{c}$) having a distance of $|\textbf{a} + \textbf{b} + \textbf{c}|/2$ between them. As a result, the phase difference between the 2$^\mathrm{nd}$ neighbors along the ($\textbf{a} + \textbf{b} + \textbf{c}$) direction with a distance of $|\textbf{a} + \textbf{b} + \textbf{c}|$ between them should be $\theta$. Hence, in a spiral state described by IP modulation vectors $\textbf{Q}_{100}$ or $\textbf{Q}_{010}$ or both, with a corresponding wavelength of $\lambda$, there shall be another modulation vector $\textbf{Q}_{111}$ along ($\textbf{a} + \textbf{b} + \textbf{c}$) with the same wavelength $\lambda$ but in units of ($\textbf{a} + \textbf{b} + \textbf{c}$). The presence of $\textbf{Q}_{111}$ as a spiral modulation vector is also evident in the FT $J(q)$ along the $G-A$ direction shown in Fig.~S4 (d). $G-A$ is near $\parallel$ to the [111] direction and $J(q)$ shows a spiral stability.  However, the confirmation comes from our ASD simulations, as discussed in the next section. Note that $\textbf{Q}_{111}$ cannot be ignored and defines the magnetic phases in GdRu$_2$Si$_2$ together with $\textbf{Q}_{100}$ and $\textbf{Q}_{010}$. This happens because of the 3D nature of the magnetic properties of the system, which must be taken into account when doing the ASD simulations. If we consider a simulation cell with the generalized dimension $N_1 \times N_2 \times N_3$, then the presence of two in-plane vectors $\textbf{Q}_{100}$ and $\textbf{Q}_{010}$ requires (i) $N_1 = N_2 = m$ where $m$ is some integer. However, the presence of $\textbf{Q}_{111}$ along with $\textbf{Q}_{100}$ and $\textbf{Q}_{010}$ requires (ii) $N_1 = N_1 = N_3 = m$ for GdRu$_2$Si$_2$ as an essential condition. Deviating from the second condition will make the ASD simulation fail to stabilize a proper spin spiral characterized by the IP vectors $\textbf{Q}_{100}$ and $\textbf{Q}_{010}$. 

\subsubsection{ASD simulations to confirm the presence of $\textbf{Q}_{111}$}

Before presenting the results from our ASD simulations to confirm the presence of $\textbf{Q}_{111}$ and the 3D nature of magnetism in GdRu$_2$Si$_2$, we briefly want to discuss some technical aspects of accurately performing these simulations. This approach to ASD simulation can be applied to many other noncollinear magnets with spiral or skyrmion phases and could be very helpful. From our calculated exchange, we get a modulation vector of $\textbf{Q}_{100}$ = (0.19, 0, 0) in units of $|\textbf{a}^*| = 2\pi/a$. This suggests an in-plane periodicity or wavelength of $\lambda_{100} = 5.26 a$ along $\textbf{a}$ corresponding to the resulting exchange spiral, which is incommensurate with the lattice periodicity. However, an integer multiple of $\lambda_{100}$, $(n\times\lambda_{100})$ could become close to $(m \times a)$ and be almost lattice-commensurate for some values of $n$, where $n$ and $m$ are integers. This information is important for choosing the size of the simulation cell ($m\times m\times m$) in ASD calculation and is provided in Table~\ref{Table1} for reference. From Table~\ref{Table1}, we can see that any arbitrary value of $m$ will not be able to accommodate an integer number of wavelengths of spiral states. As a result, for the best outcome from the ASD simulation, the values of $m$ reported in Table~\ref{Table1}, and as large as possible, shall be considered to satisfy the dimensional condition $N_1 = N_1 = N_3 = m$ for GdRu$_2$Si$_2$. Among these ``recommended'' integer $m$ values, some of them might be preferable, based on how well the condition $(n\times\lambda_{100}) \approx (m\times a)$ is satisfied. This can be characterized by the mismatch $\cfrac{n\lambda_{100} - m\cdot a}{n\lambda_{100}}\times 100\%$, which is also shown in Table~I for different $m$ values. These considerations become crucial for this system, as the exchange interactions are quite weak and frustrated. A larger deviation from the calculated values of $m$ acts as a perturbation and can distort the spiral states. Such a distortion due to unpreferred values of $m$ is, however, different from the major destabilization resulting from violating the essential condition $N_1 = N_1 = N_3 = m$ for GdRu$_2$Si$_2$. To check if our reasoning is correct, we performed ASD simulations with three different simulation cells having dimensions $37 \times 37 \times 37$, $37 \times 37 \times 21$, and $37 \times 37 \times 3$ respectively at $H = 0$ T and $T = 0$ K. Note that in these simulations we modify only one dimension of the simulation cell, while the rest of the computational parameters and the spin Hamiltonian remain the same. The first cell satisfies condition $N_1 = N_1 = N_3 = m$ with $m=37$ from Table~I and should stabilize a pure (defect-free) spiral state. In contrast, the second and third are expected to show additional deformations as we deviate from the $N_1 = N_2 = N_3 = m$ condition. In the second calculation, we do this by choosing $N_3 = 21$, which is also a good number from Table~\ref{Table1} to accommodate an integer number of spiral wavelengths but is $<$ $N_1$, $N_2$. In the third calculation, we set $N_3 = 3$, which is less than 5, the minimum value required to accommodate a single wavelength. In this case, we expect a complete failure of the ASD simulation to stabilize a proper spiral state. In these ASD simulations, the exchange interactions ($J_{ij}$) and a uniaxial anisotropy energy ($K_\mathrm{U}$) of $\sim$ 0.05 meV along the out-of-place $\textbf{c}$-axis were considered in the spin Hamiltonian (see Eqn.~\ref{eqn5}). $K_\mathrm{U}$ was calculated from DFT following a method described in detail in the methodology section. The value is in agreement with the one reported in recent literature~\cite{TNomoto_JAP.133_2023}.

\setlength{\tabcolsep}{9pt}
\begin{table}[]
\caption{Integer ($n$ = 1 to 10) multiple of the theoretically obtained in-plane spiral wavelength $\lambda_{100}$ and the closest simulation cell dimensions ($m \times m \times m$) that can accommodate this spin spiral. A mismatch between the cell dimensions and the spiral wavelength is shown as well, calculated as $\cfrac{n\lambda_{100} - m\cdot a}{n\lambda_{100}}\times 100\%$, to characterize how optimal the choice of the simulation cell is.
}

\label{Table1}
\begin{tabular}{|c|c|c|c|}
\hline
\multicolumn{4}{|c|}{${Q}_{100} = (0.19, 0, 0)\frac{2\pi}{a}$ ; $\lambda_{100}$ = 5.26a}         \\ \hline
\multicolumn{1}{|c|}{n} & \multicolumn{1}{c|}{$n\times \frac{\lambda_{100}}{a}$} & integer `m' & mismatch \\ \hline
  1  & \multicolumn{1}{c|}{5.26}  & 5   & 4.9\% \\ \hline
  2  & \multicolumn{1}{c|}{10.52} & 11  & 4.6\% \\ \hline
  3  & \multicolumn{1}{c|}{15.78} & 16  & 1.4\% \\ \hline
  4  & \multicolumn{1}{c|}{21.04} & 21  & 0.2\% \\ \hline
  5  & \multicolumn{1}{c|}{26.30} & 26  & 1.1\% \\ \hline
  6  & \multicolumn{1}{c|}{31.56} & 32  & 1.4\% \\ \hline
  7  & \multicolumn{1}{c|}{36.82} & 37  & 0.5\% \\ \hline
  8  & \multicolumn{1}{c|}{42.08} & 42  & 0.2\% \\ \hline
  9  & \multicolumn{1}{c|}{47.34} & 47  & 0.7\% \\ \hline
  10 & \multicolumn{1}{c|}{52.60} & 53  & 0.8\% \\ \hline
\end{tabular}
\end{table}
\FloatBarrier

\begin{figure}[ht]
\includegraphics[scale=0.55]{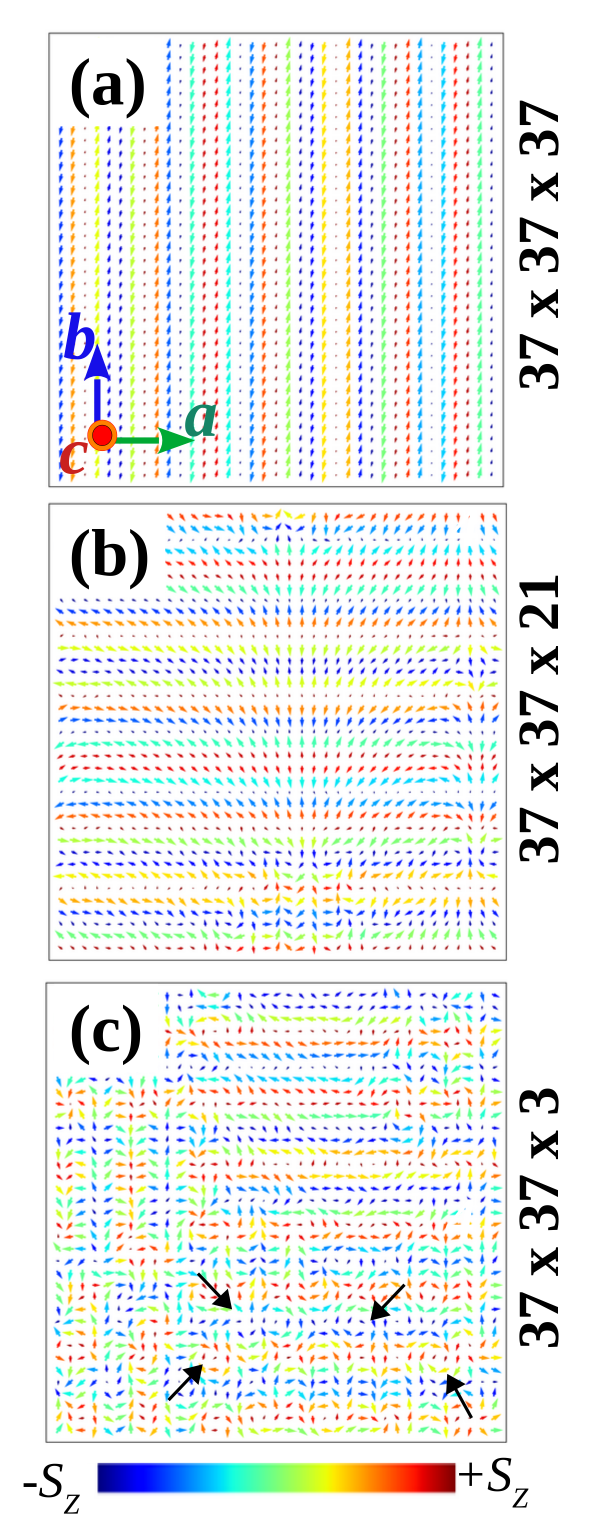}
\caption{Spin configuration after ASD simulations, of the top Gd layer of the simulation cell with dimension (a) $37 \times 37 \times 37$, (b) $37 \times 37 \times 21$, and (c) $37 \times 37 \times 3$, respectively. These ASD simulations were done using a spin Hamiltonian with only exchange ($J_{ij}$) and uniaxial anisotropy ($K_\mathrm{U}$). }
\label{dimension-distort}
\end{figure}
\FloatBarrier

\begin{figure}[ht]
\includegraphics[scale=0.55]{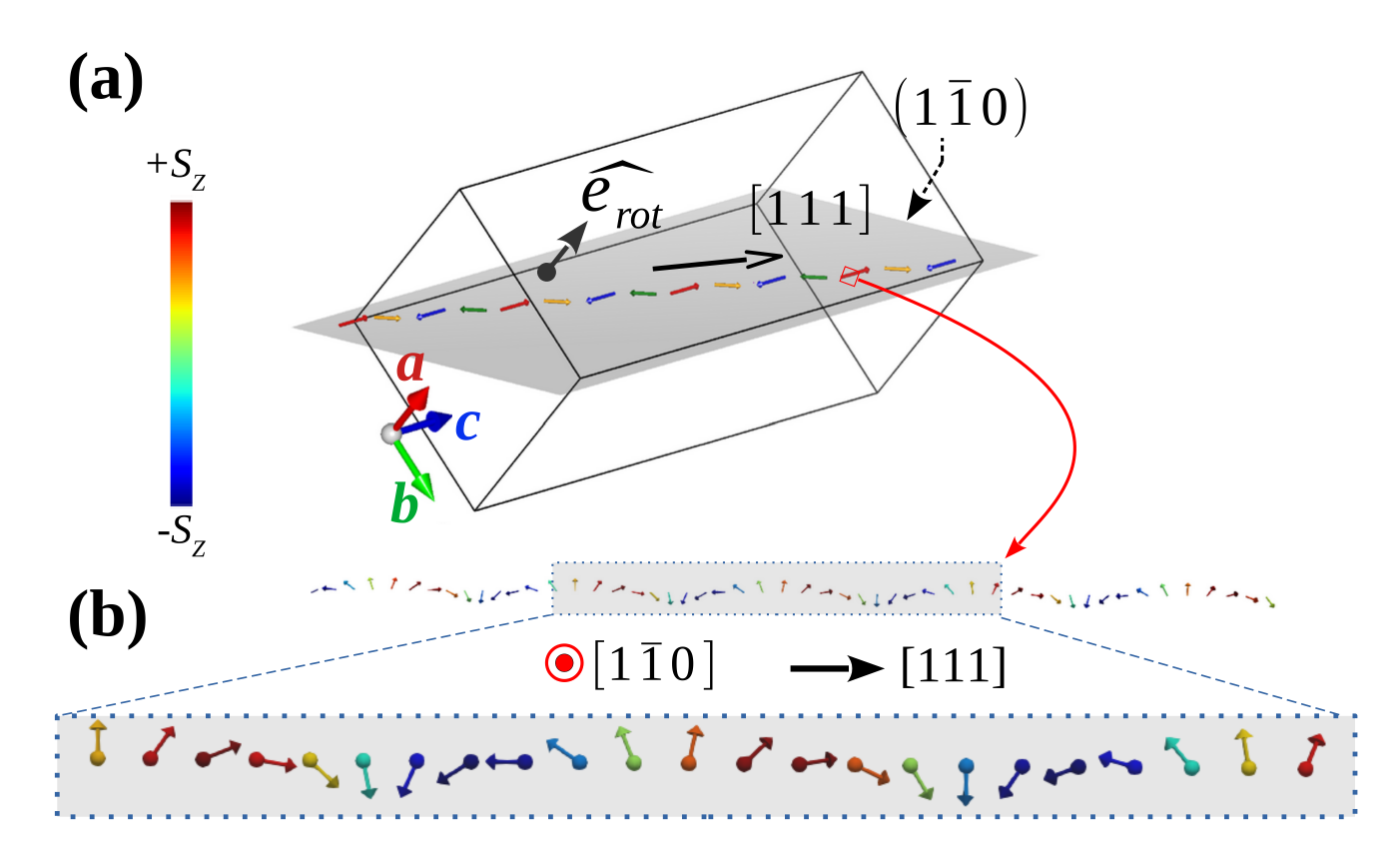}
\caption{(a) The presence of a spin-spiral in the atomic chain along the [111] direction in the spiral state where the spins in the 2D (001) planes form a helix along the [100] direction (see Fig.~\ref{dimension-distort} (a)). (b) Cycloidal nature of the interlayer spiral along $[111]$ with a spiral rotation vector along $[1\bar{1}0]$.
}
\label{interlayer-spiral}
\end{figure}

The spin configuration of the upper Gd layer of the simulation cell from the three calculations is presented in Fig.~\ref{dimension-distort}. The simulation cell of dimension $37 \times 37 \times 37$ can stabilize a defect-free helical spiral with $\textbf{Q}_{100}$ as shown in Fig.~\ref{dimension-distort} (a). This helical phase was initially reported as the ground state of GdRu$_2$Si$_2$\cite{khanh2020nanometric} but more recent experiments\cite{NDKhanh_AdvSci.9_2022,GWood_PRB.107_2023,JSpethmann_PRM.8_2024} have found a different spiral state with a double-$\textbf{Q}$ modulation.  This discrepancy could be ignored at the moment as the primary purpose of our current calculation is to obtain a defect-free spiral state resulting mainly from exchange frustration. The nature of the spiral will change based on additional interactions, e.g. magnetocrystalline anisotropy, in the Hamiltonian. The stabilization of the correct ground state will be addressed in more detail in the next section. The $37 \times 37 \times 21$ simulation cell still produces a helical state defined by a $\textbf{Q}_{010}$ but with additional distortions as shown in Fig.~\ref{dimension-distort} (b). The simulation cell $37 \times 37 \times 3$ cannot reproduce the helical state or any proper spin state that can be described by the IP $\textbf{Q}$-vectors as shown in Fig.~\ref{dimension-distort} (c). Instead, we observe a mixed disordered spin state with some isolated skyrmion-like spin vortices marked by small black arrows. This result is interesting because it opens up a new avenue to stabilize a skyrmion configuration outside of the common approach of applying an external magnetic field. Hence, this observation could be exploited to stabilize skyrmion textures in magnetic films, including the skyrmion lattice. Finally, in Fig.~\ref{interlayer-spiral} we show the arrangements of the spins in the atomic chain along the ($\textbf{a} + \textbf{b} + \textbf{c}$) or $[111]$ direction in the proper helical phase coming from the $37 \times 37 \times 37$ cell simulation shown in Fig.~\ref{dimension-distort} (a). The spin arrangements clearly show a cycloidal-type spin-spiral as shown in Fig.~\ref{interlayer-spiral} (b), propagating along the $[111]$ direction, confirming the presence of the $\textbf{Q}_{111}$ modulation vector and 3D magnetism in GdRu$_2$Si$_2$. In this cycloidal spiral, the spins lie mainly in the $(1\bar{1}0)$ plane and rotate about to the spiral rotation direction $\hat{e}_\mathrm{rot}$ nearly parallel to the $[1\bar{1}0]$ direction as shown schematically in Fig.~\ref{interlayer-spiral} (a). The nature of this spiral along [111] will also change when the present helical phase of the system changes to a different phase. However, $\textbf{Q}_{111}$ and an associated spiral will always persist because of interlayer exchange coupling.

\FloatBarrier

\begin{figure}[ht]
\includegraphics[scale=0.55]{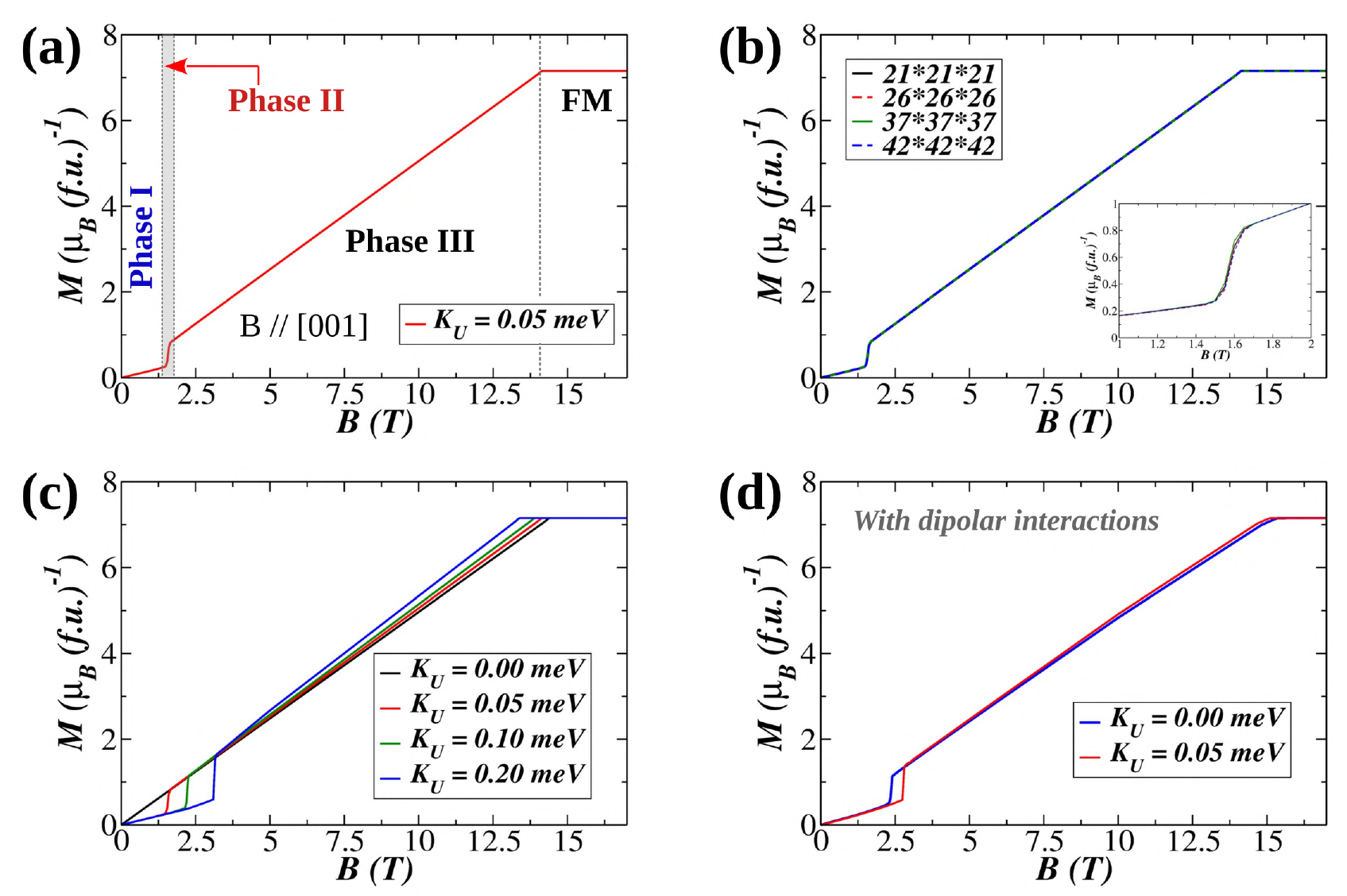}
\caption{(a) Theoretically computed $M$ vs $B$ data for $K_\mathrm{U} = 0.05$ meV and using a $37 \times 37 \times 37$ simulation cell, showing the presence of three different phases before the FM phase similar to the experimental report\cite{khanh2020nanometric}. (b) Comparison of calculated $M$ vs $B$ data with different simulation cell sizes. $K_\mathrm{U} = 0.05$ meV was used in these calculations. The inset shows a magnified version of the Phase II region. (c) $M(B)$ curves for different anisotropy constants showing the importance of $K_\mathrm{U}$. A $37 \times 37 \times 37$ simulation cell was used here. (d) $M(B)$ curves in the presence of dipolar interactions with and without anisotropy constants (where a $21 \times 21 \times 21$ simulation cell was used).}
\label{M_vs_H}
\end{figure}

To show the robustness and presence of $\textbf{Q}_{111}$ in any spiral phase, 
we show in Fig.~\ref{M_vs_H} (a) the simulated magnetic response (magnetization ($M$) vs magnetic field ($B$) measurements).  To obtain this $M(B)$ curve, a Zeeman term in the spin Hamiltonian (Eqn.~\ref{eqn5}) was considered along with exchange ($J_{ij}$) and anisotropy ($K_\mathrm{U}$).  The ASD simulations were then performed for a range of external field values applied along the $c$-axis at zero temperature with the same simulation cell $37 \times 37 \times 37$.  We obtain excellent qualitative agreement showing the presence of three distinct phases before entering the fully polarized FM phase, similar to that in the experiment\cite{khanh2020nanometric}. This establishes the physical validity of our calculated $J_{ij}$ and $K_\mathrm{U}$ on strong ground and shows the reliability of the methods used here. The $\textbf{Q}_{111}$ and interlayer spin-spin correlation also remains in $B$ induced Phase II and III.  The cycloid along [111] shown in Fig.~\ref{interlayer-spiral} from Phase I transforms into a helical and conical-type spiral in Phase II and III, respectively, with $\hat{e}_\mathrm{rot}$ near $\parallel$ [111]. A comparative view of the $\textbf{Q}_{111}$ spiral in the three phases is shown in Fig.~S5 of the SM. We end this discussion with another important observation.  The spins along the interlayer [001] direction are coupled ferromagnetically compared to the [111] direction, showing a spin spiral, as discussed above. This happens because the spins along [001] consist of the same type of atoms coupled by $J_7$, while along [111] we have two different types (see Fig.~\ref{structure} (a)) coupled by $J_2$. 

\FloatBarrier

\subsubsection{The magnetic ground-state problem}

\begin{figure}[ht]
\includegraphics[scale=0.55]{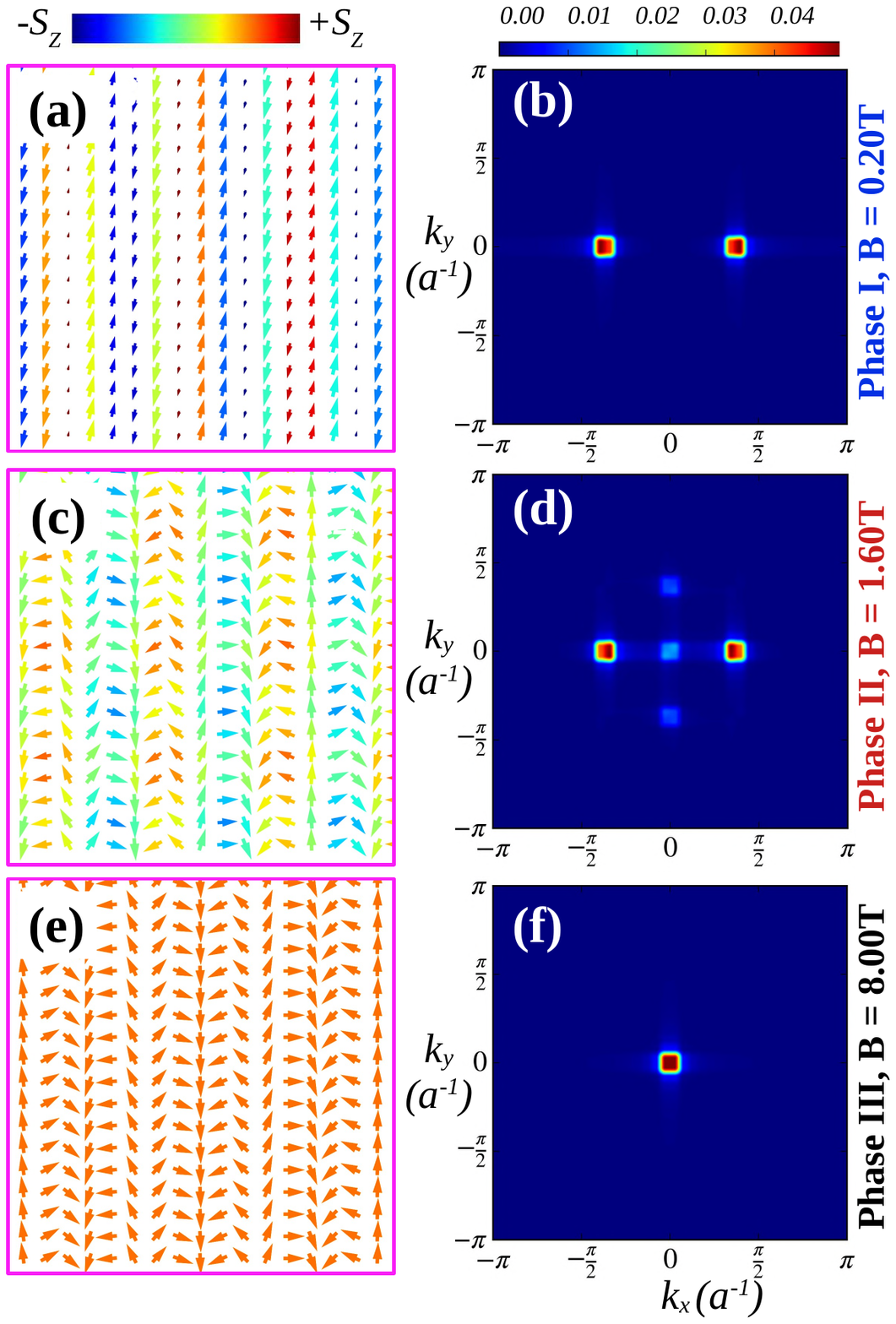}
\caption{LEFT COLUMN: Spin configuration within the 2D Gd layers in (a) Phase I, (c) Phase II, and (e) Phase III, respectively, after the ASD simulation with $K_\mathrm{U} = \unit[0.05]{meV}$, same as in Fig.~\ref{M_vs_H} (a). A $17\times17\times1$ square was cut from the $37\times37\times1$ plane for a better view of the arrangements. RIGHT COLUMN: The out-of-plane (OP) spin structure factor ($S^{OP}_s$) for (b) phase I, (d) phase II, and (f) phase III, respectively. See the text for details.}
\label{phases}
\end{figure}

This section investigates the origin of the fine modulations in the magnetic phases which require interactions beyond exchange and uniaxial anisotropy. We carefully checked the role of (i) simulation cell size, (ii) values of the calculated exchange and uniaxial anisotropy parameters, and (iii) weak magnetic interactions.  We show that reproducing the correct magnetic state/spin configuration in each phase requires accurate modeling of the system with weak magnetic interactions. Knowing the zero-field ground state is essential for correctly simulating any temperature/field-induced spin state. For example, although the spin configurations in our simulated Phases I, II, and III from Fig.~\ref{M_vs_H} (a) are different from each other, they are not the ones experimentally observed. This is not surprising because our simulation, as discussed earlier, failed to stabilize the correct ground state. Fig.~\ref{phases} (a), (c), and (e) show the spin configurations of Phases I, II, and III, respectively, from our ASD simulations. The difference could easily be verified with the $\textbf{Q}$ vectors in the calculated spin structure factor, the methodology of which is described in section~VI of the SM. The out-of-plane (OP) spin-structure factor corresponding to the three phases is shown in Figs.~\ref{phases} (b), (d), and (f), respectively. The in-plane (IP) spin structure factor of all three phases is identical, showing a single-$\textbf{Q}$ modulation, and is not shown in the main text (see Fig.~S6 of the SM for the IP structure factors).  Phase I is the helical state (similar to our zero-field ground state) with a single $\textbf{Q}$ modulation and is not the correct one according to recent experiments. This state was initially reported as the zero-field ground state by N. D. Khanh \textit{ et al.}\cite{khanh2020nanometric}. However, a double-$\textbf{Q}$ state was later identified from resonant X-ray scattering experiments with
polarization analysis\cite{NDKhanh_AdvSci.9_2022}. The spin configuration of this state was represented by an orthogonal superposition of a proper screw spiral and a sinusoidal wave\cite{NDKhanh_AdvSci.9_2022,hayami2021square}. The IP spin structure factor of this state shows a double-$\textbf{Q}$ modulation with different intensities that leads to a periodic array of vortices\cite{hayami2021square}. However, the OP structure factor shows a single-$\textbf{Q}$ modulation. The double-$\textbf{Q}$ ground state of GdRu$_2$Si$_2$ was also recently confirmed by neutron diffraction measurement \cite{GWood_PRB.107_2023} and spin-polarized scanning tunneling microscopy\cite{JSpethmann_PRM.8_2024}.   Returning to our results, in phase III, we get FM-ordered $S_z$ spin components and single-$\textbf{Q}$ $xy$ components, suggesting a conical spiral state, but recent experiments report a Meron-like crystal/lattice here\cite{NDKhanh_AdvSci.9_2022,GWood_PRB.107_2023}. Our calculated Phase II shows double-$\textbf{Q}$ modulated $S_z$ spin components with different intensities along two directions (Fig.~\ref{phases} (d)), but it is not the experimentally reported and most discussed SkL phase. The difference between the three phases in Fig.~\ref{phases} could also be realized more easily from the real space spin-density distribution of the $z$-component of the spin density and is discussed in section~VIII of the SM.

\FloatBarrier

Hence, we now focus on the reason that causes a destabilization of the correct zero-field double-$\textbf{Q}$  ground state in our ASD simulations. (A) First, we check if the simulation cell size $37 \times 37 \times 37$ is sufficient to capture the fine details of the magnetic configurations in each phase. For this, we consider a larger cell $42 \times 42 \times 42$ with $m=42$ from Table~\ref{Table1}. As can be seen, this also has a lower \% mismatch and is expected to give better results. We have also considered two lower-dimensional cells $21 \times 21 \times 21$ and $26 \times 26 \times 26$ with $m=21$ and $26$ respectively from Table~\ref{Table1} for a better comparison. The $M(B)$ curves obtained from the ASD simulations using simulation cells of different sizes are shown in Fig.~\ref{M_vs_H} (b). The same computational settings and spin Hamiltonian were used in these calculations that we used before for $37 \times 37 \times 37$. We see an excellent outcome with all the curves overlapping with each other. For a better view, a magnification of the transition regions around Phase II is shown in the inset of Fig.~\ref{M_vs_H} (b). We also confirm that the spin configurations in the three phases and the ground state of the four $M(B)$ curves in Fig.~\ref{M_vs_H} (b) are similar. We have shown these states of the $42 \times 42 \times 42$ cell in Fig.~S8 of the SM to compare with the $37 \times 37 \times 37$ ones shown in Fig.~\ref{phases}. So, the values of $m \ge 20 $ in Table~\ref{Table1} are almost equally good for a simulation cell and do not lead to incorrect spin configurations. (B) Second, we review exchange interactions from an accuracy point of view. As discussed in the last section, the soundness of our exchange parameters is undeniable from their efficiency in describing the magnetic phase transitions in Fig.~\ref{M_vs_H}(a). However, all methodologies used to calculate exchange interactions will have some inaccuracies. For example, the values of the exchange parameters calculated using a different method or a different software could differ slightly  from the ones we have (as shown in Fig.~\ref{magnetic} (a)). To check if this has anything to do with the magnetic ground state, we calculated the first three exchange parameters $J_1$, $J_2$, and $J_3$ using the total energy method. A detailed description of this method and the calculation process is given in section~X of the SM. RSPt, as well as VASP, was used for this purpose, and the calculated values are shown in Table~\ref{total-energy-j} comparatively along with the MFT calculated values from Fig.~\ref{magnetic} (a).

\begin{table}[]
\caption{Calculated values of the exchange parameters $J_1$, $J_2$, and $J_3$ using the total energy method in RSPt and VASP. The MFT calculated values from Fig.~\ref{magnetic} (a) are also tabulated for comparison purposes. }
\label{total-energy-j}
\begin{tabular}{|c|ccc|}
\hline
\multirow{2}{*}{Calculation method} & \multicolumn{3}{c|}{\begin{tabular}[c]{@{}c@{}}Exchange parameter values\\ (mRy)\end{tabular}} \\ \cline{2-4} 
 & \multicolumn{1}{c|}{$J_1$} & \multicolumn{1}{c|}{$J_2$} & $J_3$ \\ \hline
Total energy method in VASP & \multicolumn{1}{c|}{0.025} & \multicolumn{1}{c|}{0.086} & -0.009 \\ \hline
Total energy method in RSPt & \multicolumn{1}{c|}{0.018} & \multicolumn{1}{c|}{0.081} & -0.011 \\ \hline
Magnetic Force Theorem in RSPt & \multicolumn{1}{c|}{-0.004} & \multicolumn{1}{c|}{0.095} & -0.012 \\ \hline
\end{tabular}
\end{table}
\FloatBarrier

We get a good agreement between the RSPt and VASP total energy calculations, which is a good sign since we are using the same method here. However, these values do not fully agree with the MFT-calculated values of $J_1$, $J_2$, and $J_3$. Although $J_2$ and $J_3$ are all very similar, $J_1$ in the total energy calculations is noticeably larger and also FM in nature compared to its weak AFM counterpart from the MFT calculation. This could indeed alter the nature of the zero-field ground state and this was investigated. Since the total energy calculated values from RSPt and VASP are very close, only the VASP calculated values were considered for this check. First, to check the dominant $\textbf{Q}$ vectors in this case, the MFT calculated $J_1$, $J_2$, and $J_3$ in our $J_{ij}$s were replaced with the VASP calculated values from Table~\ref{total-energy-j}.  Then FT $J(q)$ was calculated for $\textbf{q}$ $\parallel$ different directions of the Brillouin zone and are shown in Fig.~S10 of the SM. The results are qualitatively the same as before with $\textbf{Q}_{100}$ and $\textbf{Q}_{010}$ showing maximum stability over other spiral modulation vectors. In this case, we get  $|\textbf{Q}_{100}|$ =  $|\textbf{Q}_{010}|$ = 0.18$\textbf{a}^*$, slightly less than 0.19$\textbf{a}^*$ obtained before with all MFT calculated values of $J_{ij}$s. Hence, we can expect a very similar helical spiral ground state from here also. To confirm this, a $(B,T) = 0$ ASD simulation was performed with this new set of exchange parameters and $K_U = 0.05$ meV. The helical spin state in the 2D Gd layer and its characterization from the IP and OP spin structure factor are shown in Fig.~S11 of the SM. This analysis suggests that as long as we obtain the stabilization of $\textbf{Q}_{100}$ and $\textbf{Q}_{010}$ from some exchange parameters ($J_{ij}$), an ASD simulation along with some $K_\mathrm{U}$ would possibly stabilize some single-$\textbf{Q}$ ground state. Small deviations/inaccuracies in calculated $J_{ij}$s that come from different calculation methods should not make any difference and cannot be a reason for missing the correct double-$\textbf{Q}$ ground state. This statement is also supported by the data presented in a recent work by Bouaziz \textit{et al.} \cite{bouaziz2022fermi} where ASD simulation with their $J_{ij}$s showing the same $\textbf{Q}_{100}$ stability and $K_\mathrm{U}$ resulted in the same helical ground state that is presented here. Hence, for any further calculations and analysis, we stick to our MFT calculated exchange parameters in Fig.~\ref{magnetic}(a). (C) Third, we have to discuss the role of uniaxial anisotropy $K_\mathrm{U}$. As our ASD simulations are using the DFT calculated anisotropy value of $K_\mathrm{U}$ = 0.05 meV, a number that might have uncertainty in it, the influence of the magnetic anisotropy on the magnetic configuration also needs to be checked. The simulated $M(B)$ curves for different values of $K_\mathrm{U}$ are shown in Fig.~\ref{M_vs_H}(c). For these calculations, we used the same $37 \times 37 \times 37$ simulation cell, and all other computational settings were kept fixed. An increased anisotropy value of 0.10 meV gives a much better description of the magnetic response (green curve). Phase II now appears around the experimental value of $B = 2.00$ T compared to the $K_\mathrm{U}$ = 0.05 meV case that shows it around 1.50 T. However, even with this improvement, the spin configurations in each phase, including the zero-field ground state, do not change. All come out to be the same as before, along with the same helical state at zero-field. The spin configurations and the corresponding spin structure factors of these phases are shown in Fig.~S12 of the SM. For a deeper understanding of the role of $K_\mathrm{U}$, it was removed from a set of ASD simulations to see the effect on the $M(B)$ curve. This gives a dramatic outcome where exchange alone fails to predict the known magnetic phase transitions (black curve). The system directly enters the field-polarized FM state from the only observed conical spiral state. This observation emphasizes the importance of $K_\mathrm{U}$ in GdRu$_2$Si$_2$. This also aligns with the study of Bouaziz \textit{et al.} \cite{bouaziz2022fermi}, where a large value of $K_\mathrm{U}$ around 0.30~meV was reported to have been used in the ASD simulations to stabilize a SkL phase under an applied magnetic field. Following reference~\cite{bouaziz2022fermi}, we further increased the value of $K_\mathrm{U}$ to 0.20 meV. This affected the $M(B)$ curve, pushing Phase II even higher along the $B$-axis, but the nature of the phases did not change (data not shown here).  With larger $K_\mathrm{U}$ values, Bouaziz \textit{et al.} \cite{bouaziz2022fermi} also reports the same zero-field helical spiral ground state and a high-field conical spiral state similar to our results, contradicting the recent experimental data\cite{NDKhanh_AdvSci.9_2022,GWood_PRB.107_2023,JSpethmann_PRM.8_2024}. However, their calculation surprisingly shows a SkL state for some critical values of the $B$-field. One possible reason for this stabilization could be the $60 \times 60 \times 6$ simulation cell used in their calculation, deviating from the $N_1=N_2=N_3$ condition we propose for this system. We have already shown in the previous section (Fig.~\ref{dimension-distort} (c)) that such a cell could induce skyrmion-like objects even at $B = \unit[0]{T}$. Hence, a simulation cell of this kind, along with $K_\mathrm{U}$ around 0.30~meV, could stabilize a SkL state as was claimed~\cite{bouaziz2022fermi}. However, considering the 3D nature of magnetism in GdRu$_2$Si$_2$, the SkL state reported in Ref.\cite{bouaziz2022fermi} could be a special case. Hence, from here we can conclude that $K_\mathrm{U}$ is essential to correctly describe the magnetic phase transitions of GdRu$_2$Si$_2$, but not a sufficient element to stabilize the correct spin states or the double-$\textbf{Q}$ ground state.

\begin{figure}[ht]
\includegraphics[scale=0.55]{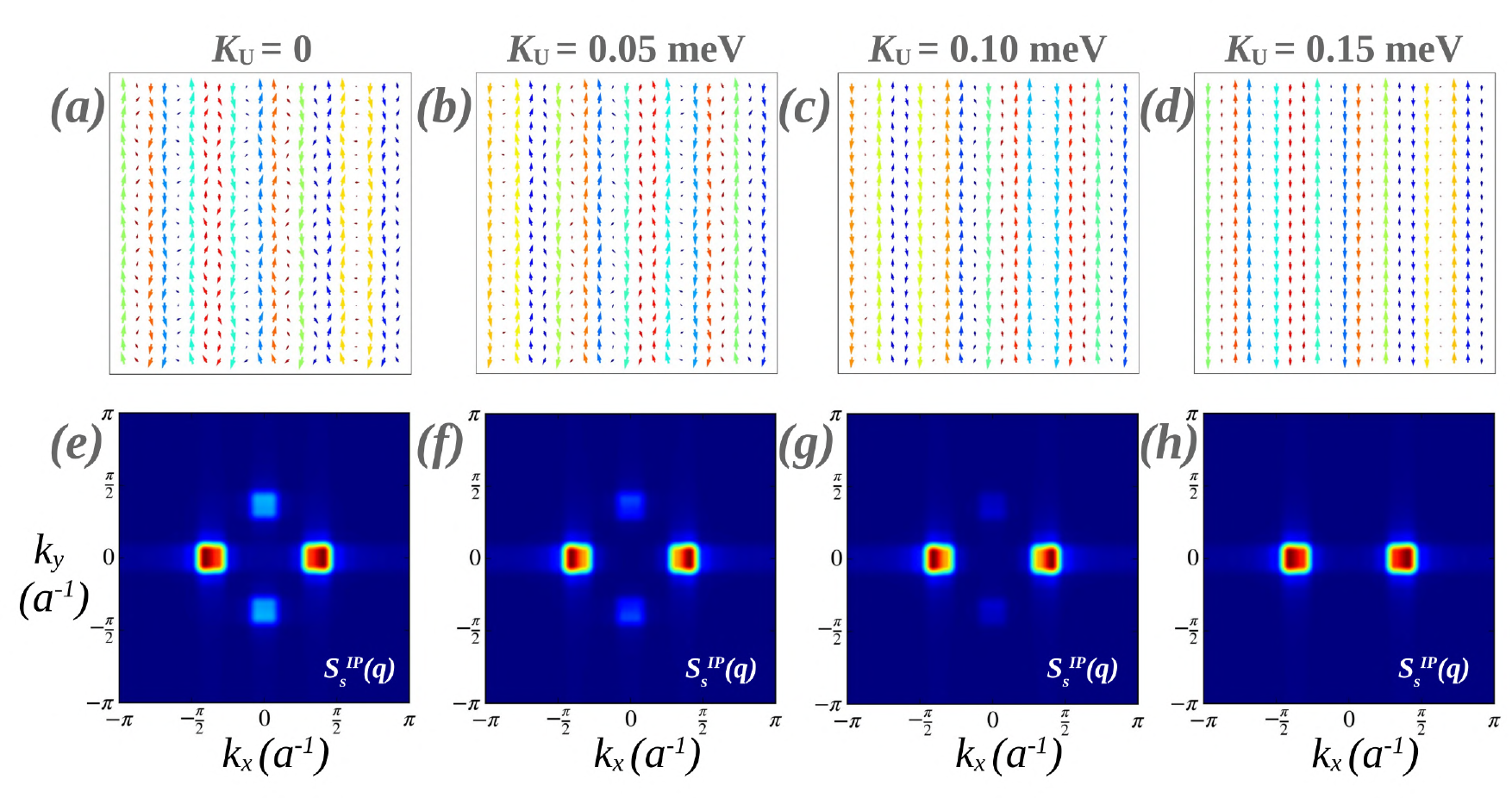}
\caption{Effect of uniaxial anisotropy $K_\mathrm{U}$ in the presence of dipolar interactions. TOP PANEL: Spin configuration within the $21\times21\times1$ 2D Gd layers for (a) $K_\mathrm{U} = 0$, (b) $K_\mathrm{U} = 0.05$,  (c) $K_\mathrm{U} = 0.10$ and (d) $K_\mathrm{U} = 0.15$ meV respectively. BOTTOM PANEL: The corresponding IP spin structure factor ($S^{IP}_s$)  of the spin states shown in the top panel. See the text for more details.}
\label{dip-role}
\end{figure}
\FloatBarrier

\begin{figure}[ht]
\includegraphics[scale=0.55]{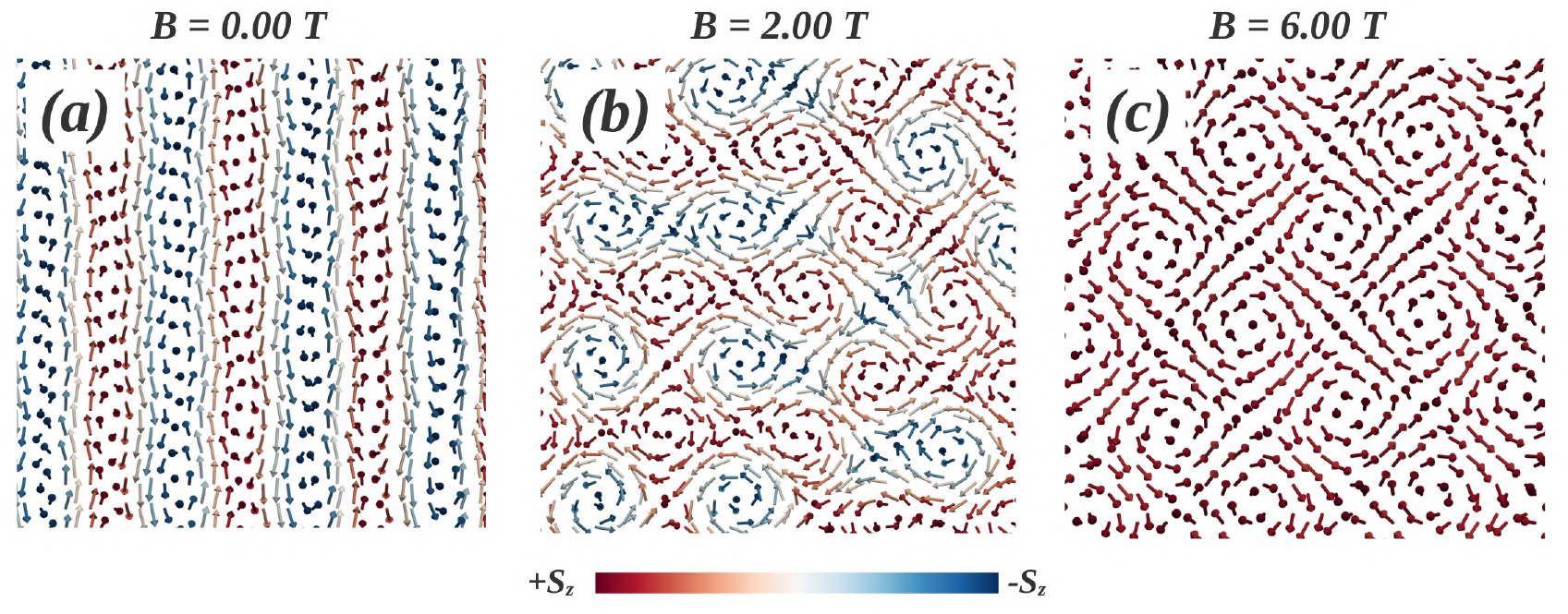}
\caption{Spin configuration within the 2D Gd layers after ASD simulation from a $60\times60\times10$ cell without periodic boundary conditions at (a) $B =0.0$ T, (b) $B =2.0$ T, and (c) $B =6.0$ T, respectively.  A Spin Hamiltonian with the same exchange parameters shown in Fig.~\ref{magnetic} (a), dipolar interactions, and $K_\mathrm{U} = \unit[0.05]{meV}$ was considered. A $20\times20\times1$ square was cut from the $60\times60\times1$ plane for a better view of the spin arrangements.}
\label{606010}
\end{figure}

(D) Lastly, we want to discuss the role of weak magnetic interactions not considered in our spin Hamiltonian. The microscopic magnetic interactions we have been considering are the isotropic Heisenberg exchange (bilinear) and uniaxial anisotropy. Although the on-site uniaxial anisotropy ($K_\mathrm{U}$) cannot change the single-$\textbf{Q}$ helical ground state, some other inter-site interactions like exchange could. In this regard, weak magnetic interactions such as the biquadratic interaction, dipolar interaction, and symmetric anisotropic exchange (SAE) interaction could play an important role. A complete understanding will need to consider each of these weak interactions separately and their all possible combinations with exchange ($J_{ij}$) and anisotropy ($K_\mathrm{U}$). This is going to be a separate study on its own and is beyond the scope of our present work. However, among the three interactions mentioned, the magnetic dipolar interaction should not be ignored. Magnetic dipolar interactions are naturally present in any magnetic material and could become important in a system like GdRu$_2$Si$_2$, where large atomic moments of $\unit[7.0]{\mu_B}$ are present, potentially leading to non-negligible dipolar interactions especially compared to relatively small Heisenberg interactions (Fig.~\ref{magnetic} (a)). The role of dipolar exchange in stabilizing topological magnetic structures like magnetic bubbles or skyrmions is well known from previous studies of similar systems~\cite{YSLin_APL.23_1973,TGarel_PRB.26_1982,STakao_JMMM.31_1983}. Hence, we now add the dipolar interaction term defined in Eqn.~\ref{eqn6} within our spin Hamiltonian in Eqn.~\ref{eqn5} with exchange, uniaxial anisotropy, and the Zeeman term for ASD simulations. The inclusion of dipolar interactions makes the ASD simulations computationally very demanding. 
To reduce the computational load, we used the simulation cell $21\times21\times21$, with a tetragonal shape, which is equally accurate as the $37\times37\times37$ cell, as discussed at the very beginning of this section (Fig.~\ref{M_vs_H} (b)).  In the previous paragraph, we have also discussed the importance of on-site uniaxial anisotropy ($K_\mathrm{U}$) for a correct description of the magnetic phase transitions in GdRu$_2$Si$_2$ (Fig.~\ref{M_vs_H} (c)). For this reason, we perform $(B,T) = 0$ ASD simulations with dipolar interactions and for different values of $K_\mathrm{U}$ to check the possible ground-state structures.  The spin configurations in the 2D Gd layers of the resulting ground states are shown in the top panel of Fig.~\ref{dip-role}, and the bottom panel shows the corresponding IP spin structure factors. The OP spin structure factor for all the cases shows similar single-$\textbf{Q}$ modulation of the spins and is not shown here in the main text (see Fig.~S13 of the SM for the OP case). For $K_\mathrm{U}$ = 0.15 meV, we get a perfect helical spiral with a single-$\textbf{Q}$ modulation. As the value of $K_\mathrm{U}$ is decreased, the spins start to show some planar rotations, which become very pronounced in the form of some vortices when $K_\mathrm{U}$ is removed. This is not a helical spiral state, and it is clear from the IP spin structure factor showing a double-$\textbf{Q}$ modulation with different intensities along the two in-plane directions. Such rotational behavior of the in-plane components of the spins and the corresponding double-$\textbf{Q}$ structure factor in Fig.~\ref{dip-role} (a) and (e), respectively, matches with the experimentally reported double-$\textbf{Q}$ ground state\cite{NDKhanh_AdvSci.9_2022, GWood_PRB.107_2023, JSpethmann_PRM.8_2024}.  These results show the importance of dipolar interactions for the correct ground state properties of the system. This spiral state persists for our DFT calculated value of $K_\mathrm{U} = 0.05$ meV but with a lower intensity and almost dissipates above this value. For  $K_\mathrm{U} = 0.15$ meV, on-site uniaxial anisotropy dominates over dipolar interactions, stabilizing the helical spiral state again. This observation is, however, not surprising but more realistic because a small value of  $K_\mathrm{U}$ is something to expect in GdRu$_2$Si$_2$. In GdRu$_2$Si$_2$, Gd$^{3+}$ cations with $S = 7/2$ and $L = 0$ are expected to show a very small or zero value of $K_\mathrm{U}$ due to the zero orbital moment in this case~\cite{AGarnier_JMMM.140_1995}. To understand the importance of $K_\mathrm{U}$ in the presence of dipolar interactions, we again calculate the $M(B)$ curve without any $K_\mathrm{U}$ (blue line) as shown in Fig.~\ref{M_vs_H} (d). Interestingly, we get an excellent qualitative description of the magnetic transitions showing all three phases even without $K_\mathrm{U}$. Phase II also appears around the experimental value of $B= 2.0~T$. This is strikingly different from the data without dipolar interactions (see Fig.~\ref{M_vs_H} (c)), where  $K_\mathrm{U}$ was essential to simulate the magnetic response of the system correctly.  Since we get the correct ground state and magnetic transitions with dipolar interactions and without any $K_\mathrm{U}$, we can anticipate the experimentally reported SkL state in Phase II and a Meron-like state in Phase III. The spin configurations in the three phases and their characterization with the spin structure factor are given in Fig.~S14 of the SM. Phases II and III do not show a SkL and Meron-like lattice but are very similar to what we got before without dipolar interactions (Fig.~\ref{phases}).  One reason behind the absence of a SkL and Meron-like lattice in Phase II and III could be the exclusion of $K_\mathrm{U}$ from these calculations. Although in the presence of dipolar interactions $K_\mathrm{U}$ looks a bit irrelevant, a small value could be important for any $B$-induced spin states.  To check this, the $M(B)$ curve was calculated with our DFT calculated value of $K_\mathrm{U} = 0.05$ meV (red line) and is also shown in Fig.~\ref{M_vs_H} (d).  Phase II is again pushed higher in the $B$-axis as expected but no other changes in the spin configurations of any phases were observed. We get the same spin states in the three phases observed earlier for the $K_\mathrm{U} = 0$ case (data not shown here). From Fig.~\ref{dip-role}, we already know that a larger value of $K_\mathrm{U}$ destabilizes the correct ground state properties and hence should not be used aiming to stabilize a SkL state in phase II or a Meron-like configuration in Phase III.  For this reason and due to the demanding nature of these dipolar calculations, we have not calculated the $M(B)$ curves for any larger $K_\mathrm{U}$ values. Our results show the importance of dipolar interactions in GdRu$_2$Si$_2$ for obtaining the correct magnetic ground state and also play a similar role like the uniaxial anisotropy to control the magnetic phase transitions. Our data also suggest that other weak interactions, like the biquadratic interaction and symmetric anisotropic exchange interaction, could be important to stabilize the SkL state in Phase II and needs to be checked carefully. For a symmetric tetragonal crystal like this, the importance of biquadratic interaction mediated by itinerant electrons in the presence of an easy axis anisotropy to stabilize a multiple-$\textbf{Q}$ SkL state over a single-$\textbf{Q}$ spiral state has been claimed before ~\cite{khanh2020nanometric,hayami2021square,NDKhanh_AdvSci.9_2022}. The theoretical work by Hayami \textit{et al.}~\cite{hayami2021square} considers a model Hamiltonian based on a Kondo lattice model consisting of itinerant electrons and localized spins. The Hamiltonian included both bilinear and biquadratic interaction, and its role was analyzed for a 2D square lattice. The biquadratic exchange interaction was found to be important in stabilizing any spin state with a double-$\textbf{Q}$ modulation, which could be even further stabilized by bond-dependent anisotropic interactions. Such states could transform into a proper SkL state with the application of an external magnetic field. But as mentioned earlier, this needs to be investigated in detail in our ASD simulations and could only be addressed in a separate study since, in this work, we have already addressed multiple important questions about the magnetic properties of GdRu$_2$Si$_2$.  To elaborate on this point, we would also like to mention that it could be possible to stabilize a SkL state in a 2D square lattice using some model Hamiltonians under an external $B$-field. However, this generally depends on a rigorous fitting of some parameters to achieve the desired results and doesn't necessarily show the reality of the system. As an example, we present in Fig.~\ref{606010} some ASD simulation results with some artificial simulation conditions. The spin Hamiltonian remains the same with exchange, uniaxial anisotropy, and dipolar interactions, but a slab-like $60\times60\times10$ simulation cell was used without any periodic boundary conditions. At $T=0$ K, the experimentally reported correct zero-field ground state and field-induced SkL at $B=2.0$ T, as well as the Meron state, could be obtained but with some small defects. Although these results align with the experimental reports, a physical explanation is not straightforward, as the simulation conditions used are not in line with the physical properties of the system.  Our primary simulations presented before, on the contrary, do not involve any fitting process or artificial manipulations. The values of the parameters used in our ASD simulations are obtained from DFT calculations and hence provide a realistic picture of the system. This, however, comes with a cost where one needs to systematically check the role of all possible physical interactions, making it time-consuming and not feasible within a single study.  Although our simulations here do not stabilize the $B$-induced, most discussed SkL state, the zero-field ground state, and magnetic transitions are correctly captured. Based on these calculations, we can reveal the correct 3D nature of magnetism in this system. By restarting from here and improving the Hamiltonian systematically, one could gradually get all the states correctly. Hence, our results are important and should be considered as the correct steps to understand GdRu$_2$Si$_2$'s magnetism from first principles.

\FloatBarrier

\section{Conclusion}

GdRu$_2$Si$_2$ is a metallic magnet with Gd $4f$ moments interacting via the RKKY exchange mechanism. This triggers an exchange frustration in the system leading to non-collinearity even without any DMI. Our calculated exchange interactions confirm the RKKY behavior, and the main source of exchange frustration is the competition between the interlayer and intralayer exchange couplings, which are dominantly FM and AFM in nature, respectively.  Interestingly, the interlayer FM exchange is much stronger compared to the intralayer AFM interactions, suggesting that GdRu$_2$Si$_2$ is a 3D bulk magnet, despite its layered-type appearance. With further analysis of the exchange data, we can identify the presence of a previously unnoticed interlayer modulation vector $\mathbf{Q}_{111}$ that coexists with the modulation vectors $\mathbf{Q}_{100}$ and $\mathbf{Q}_{010}$, known from the literature. We could confirm the $\mathbf{Q}_{111}$-modulation of the spin texture in the spiral phases from our spin-dynamics simulations and show that it is in line with its magnetic properties. This interlayer modulation vector cannot be ignored and describes the magnetic phases together with the experimentally observed intralayer modulation vectors $\mathbf{Q}_{100}$ and $\mathbf{Q}_{010}$. This makes the magnetic phases of GdRu$_2$Si$_2$ far more complex than they appear on the 2D Gd layers, where the interlayer spin correlation results in a complex modulation of the spins. By taking these correlations into account, our spin dynamics simulations can reproduce the magnetic-field-induced phase transitions, in good qualitative agreement with the experimental data. We also demonstrate the importance of dipolar interactions in governing the ground-state magnetic properties of the system. A competition between the effects of dipolar interaction and uniaxial anisotropy was also found,  and these two types of interactions attempt to stabilize different magnetic orderings. The dominance of the dipolar interaction over uniaxial anisotropy becomes evident from the ground-state properties. This indirectly points to a weak uniaxial anisotropy consistent with the spherical symmetry of the Gd $4f$ states. However, our simulations are not able to stabilize the experimentally reported SkL and Meron-like states at higher field values. We ascribe this to weak magnetic interactions such as biquadratic and symmetric anisotropic exchange that are absent in our spin Hamiltonian. Nevertheless, our observations so far represent important steps towards a complete understanding of the magnetism of GdRu$_2$Si$_2$, especially in light of DFT and ASD simulations.

Regarding the computational methodology, we point out the importance of supercell dimensions for modeling the magnetic state using atomistic spin dynamics. Quite different results can be obtained depending on the dimensions, even along the $c$-axis, relative to the wavelength of the spin spirals and sizes of the skyrmions. This can explain the difference between some of our results and previous studies. This observation can be important for spin dynamics studies of magnetic systems in general and for non-collinear magnets in particular, where the size effects in terms of the simulation cell dimensions can be of high importance, especially when magnetic textures are not commensurate with the crystal lattice.  This also carries over to the experimental investigations where thin film growth and/or nano-patterning could be used to induce and manipulate specific skyrmionic states. Further studies and experiments in this direction would potentially provide new insights and advance the field of topological magnetism.

\section{Acknowledgments}

This work was financially supported by the Knut and Alice Wallenberg Foundation through grant numbers 2018.0060, 2021.0246, and 2022.0108 (PI's: O.E. and A.D.), and R.P. and V.B. acknowledge support from the G\"oran Gustafsson Foundation (recipient of the ``small prize'': V.B.). V.B. also acknowledge support from the Ministry of Education, Youth and Sports of the Czech Republic through the e-INFRA CZ (ID:90254). O.E. and A.D. acknowledge support from the Wallenberg Initiative Materials Science for Sustainability (WISE) funded by the Knut and Alice Wallenberg Foundation (KAW). A.D. also acknowledges financial support from the Swedish Research Council (Vetenskapsrådet, VR), Grant No. 2016-05980, Grant No. 2019-05304, and Grant No. 2024-04986. O.E. also acknowledges support by the Swedish Research Council (VR), the Foundation for Strategic Research (SSF), the Swedish Energy Agency (Energimyndigheten), the European Research Council (854843-FASTCORR), eSSENCE and STandUP. S.S. acknowledges funding (postdoctoral stipend) from the Carl Tryggers Foundation (grant number CTS 22:2013, PI: V.B.).

The computations/data handling were enabled by resources provided by the Swedish National Infrastructure for Computing (SNIC) at the National Supercomputing Centre (NSC, Tetralith cluster) partially funded by the Swedish Research Council through grant agreement no.\,2018-05973 and by the National Academic Infrastructure for Supercomputing in Sweden (NAISS) at the National Supercomputing Centre (NSC, Tetralith cluster) partially funded by the Swedish Research Council through grant agreement no.\,2022-06725. We acknowledge VSB – Technical University of Ostrava, IT4Innovations National Supercomputing Center, Czech Republic, for awarding this project access to the LUMI supercomputer, owned by the EuroHPC Joint Undertaking, hosted by CSC (Finland) and the LUMI consortium through the Ministry of Education, Youth and Sports of the Czech Republic through the e-INFRA CZ (grant ID: 90254). We also acknowledge EuroHPC Joint Undertaking for awarding us access to Karolina supercomputer at IT4Innovations, Czech Republic. Structural sketches in Figs.~\ref{structure}--\ref{helical} were produced using the \textsc{VESTA3} software \cite{vesta}. Spin configurations in Figs.~\ref{dimension-distort} and \ref{phases} were plotted using the SpinView software \cite{spinview}.

We thank Saikat Sarkar (CRIS, India) for useful discussions on this study, Nastaran Salehi and Philipp Rybakov for discussions about the dipolar exchange and verification of its implementation in the UppASD software, Manuel Pereiro and Qichen Xu for similar discussions, and Anders Bergman for general discussions about the UppASD software.

\medskip
\bibliographystyle{naturemag.bst}
\bibliography{main}

\end{document}


\title{Supplementary material for: Unveiling Mysteries of GdRu$_2$Si$_2$: 3D Magnetism in a layered like Magnet}
\author{Sagar Sarkar}
\affiliation{Department of Physics and Astronomy, Uppsala University, Uppsala, 751 20, Sweden.}

\author{Rohit Pathak}
\affiliation{Department of Physics and Astronomy, Uppsala University, Uppsala, 751 20, Sweden.}

\author{Arnob Mukherjee}
\affiliation{Department of Physics and Astronomy, Uppsala University, Uppsala, 751 20, Sweden.}

\author{Anna Delin}
    \affiliation{\KTH}
    \affiliation{\WISEKTH}
    \affiliation{\SeRC}

\author{Olle Eriksson}
\affiliation{Department of Physics and Astronomy, Uppsala University, Uppsala, 751 20, Sweden.}
\affiliation{Wallenberg Initiative Materials Science for Sustainability, Uppsala University, 75121 Uppsala, Sweden.}

\author{Vladislav Borisov}
\affiliation{Department of Physics and Astronomy, Uppsala University, Uppsala, 751 20, Sweden.}


\maketitle

\section{Comparison between VASP ($\mathrm{Gd}$ $4f$ states in the valence bands) and RSPt ($\mathrm{Gd}$ $4f$ in the core) calculated electronic structure.}

\begin{figure}[ht]
\includegraphics[scale=0.5]{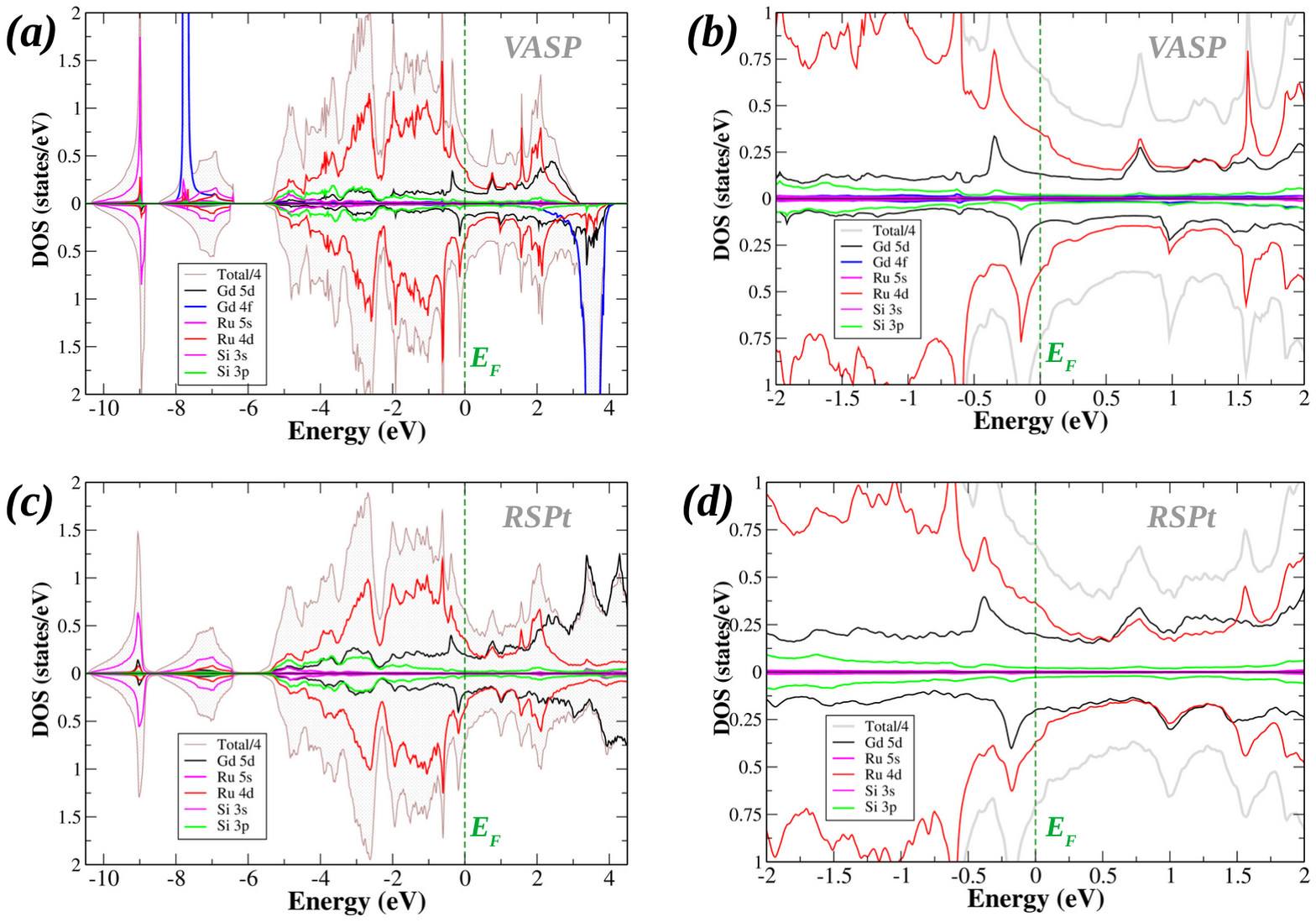}
\caption{TOP PANEL: (a) Total (scaled by 4) and atom+orbital projected density of states from VASP calculation where Gd $4f$ states were treated as valence states. (b) Zoom-in of plot (a) around the Fermi energy.  BOTTOM PANEL: (c) Total (scaled by 4) and atom+orbital projected density of states from RSPt calculation where Gd $4f$ states were in the core. (d) Zoom-in of plot (c) around the Fermi energy. The plots of RSPt data do not show the Gd $4f$ states, because they are not part of the valence states. For VASP results, the Gd $4f$ states are indicated by the blue lines.}
\label{dos_compare}
\end{figure}

A comparison between the electronic structure calculated in VASP and RSPt is shown in Fig.~\ref{dos_compare}. In the VASP calculations, the localized Gd $4f$ states were considered as part of the valence bands, and a DFT+$U$ calculation was performed (see the main text for computational details). In contrast, RSPt calculations are done assuming that the Gd $4f$ states are in the core and have zero energy dispersion. Notice that the bottom panel of Fig.~\ref{dos_compare} showing the RSPt data does not have the Gd $4f$ state (blue lines). We get a very good quantitative agreement between both codes (VASP and RSPt) despite the use of different basis sets and computational methodologies. This agreement confirms the localized non-interacting nature of the half-filled Gd $4f$ shell and supports our decision to treat them as core states while computing the inter-atomic exchange interactions (see main text for more details).

\FloatBarrier

\section{Comparison of calculated exchange interactions without and with SOC.}

\begin{figure}[ht]
\includegraphics[scale=0.5]{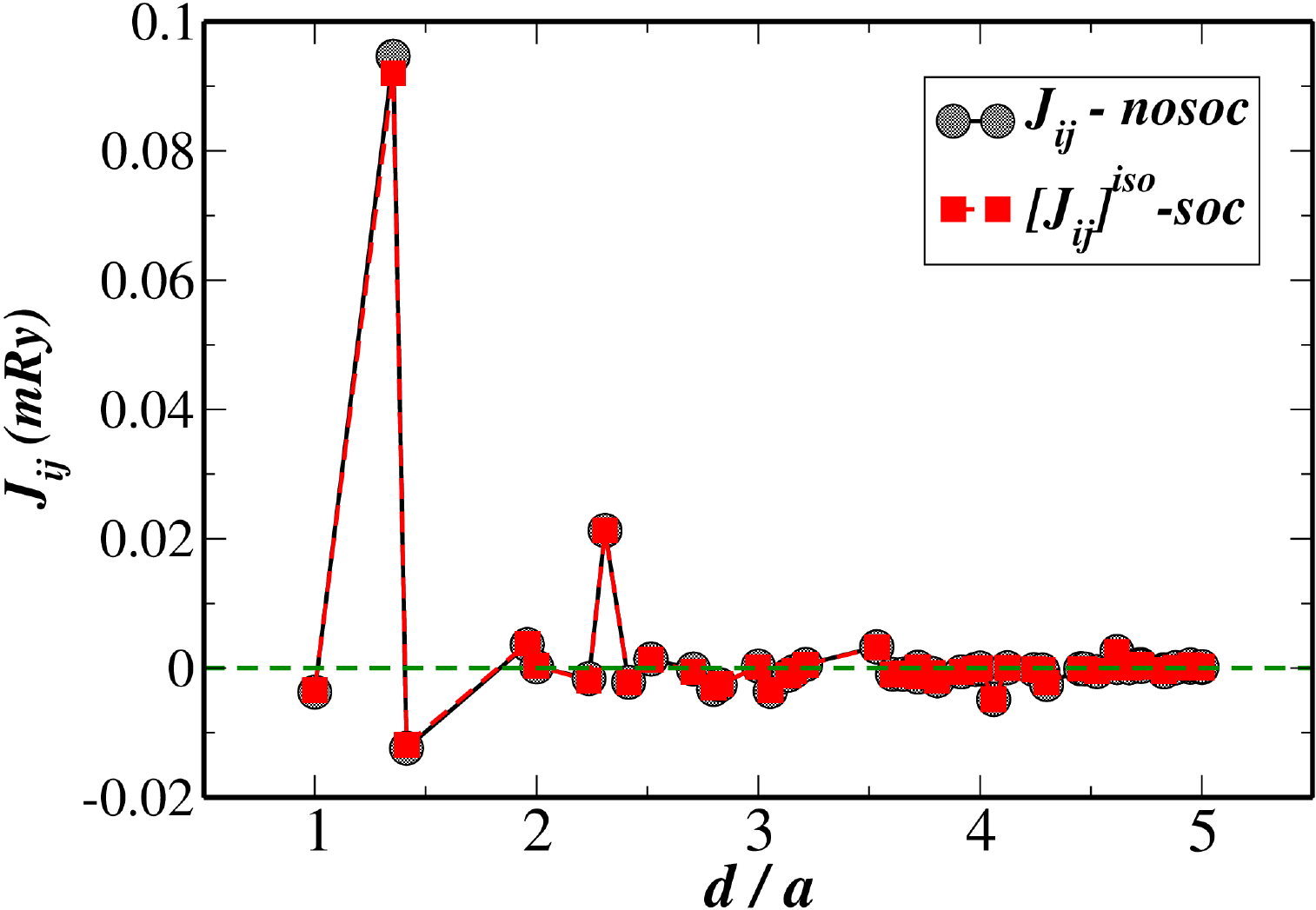}
\caption{Calculated $Gd-Gd$ exchange without (black spheres) and with (red squares) SOC. For the latter, only
the isotropic averages of the exchange tensor are plotted. }
\label{soc-nosoc}
\end{figure}

Our system has heavy Gd atoms with highly localized and anisotropic $4f$ states. As a result, the relativistic effects are important and could influence the interatomic exchange interactions. We have also calculated the $Gd-Gd$ exchange tensor with SOC to check this. Then we compared the isotropic averages of the exchange tensor with the non-relativistic exchange interactions described in the main text. This is shown in Fig.~\ref{soc-nosoc} for $R_{d/a} = 5$ (see the main text for more details). There is no major difference suggesting that relativistic effects are not very important.

\FloatBarrier

\section{High symmetry points in the reciprocal space Brillouin Zone}

\begin{figure}[ht]
\includegraphics[scale=0.5]{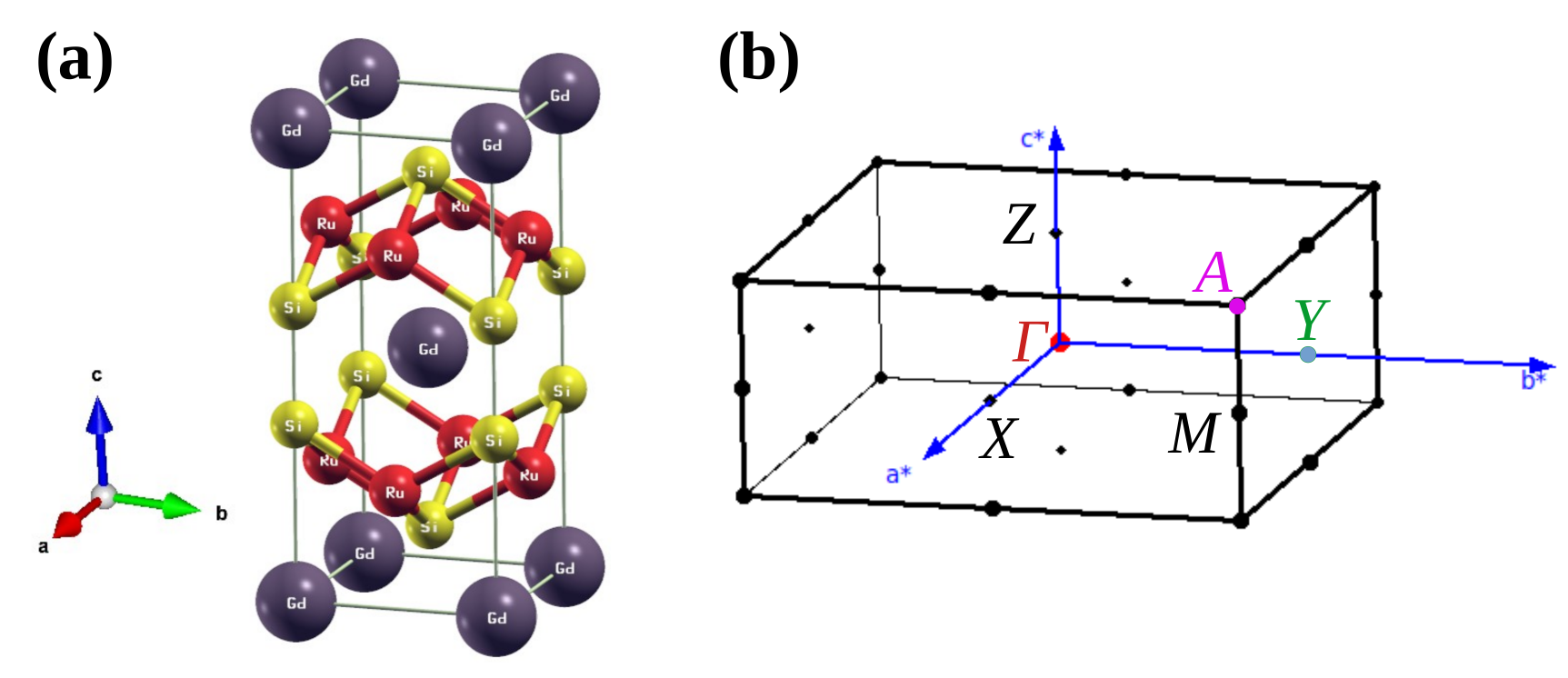}
\caption{The tetragonal GdRu$_2$Si$_2$ (a) unit cell and corresponding (b) reciprocal space Brillouin zone (BZ) showing the high-symmetry (HS) points.}
\label{bz}
\end{figure}

The tetragonal GdRu$_2$Si$_2$ unit cell and the corresponding reciprocal space Brillouin zone (BZ) with the high-symmetry (HS) points are shown in Fig.~\ref{bz}. The $\Gamma-X$, $\Gamma-Y$, $\Gamma-Z$, and $\Gamma-M$ directions are $\parallel$ to the crystallographic [100]($\textbf{a}$-axis), [010]($\textbf{b}$-axis), [001]($\textbf{c}$-axis), and [110] directions respectively. The direction $\Gamma-A$ is nearly parallel to the crystallographic direction [111].

\FloatBarrier

\section{FT $J(q)$ of exchange along high symmetry directions of the Brillouin Zone}

\begin{figure}[ht]
\includegraphics[scale=0.5]{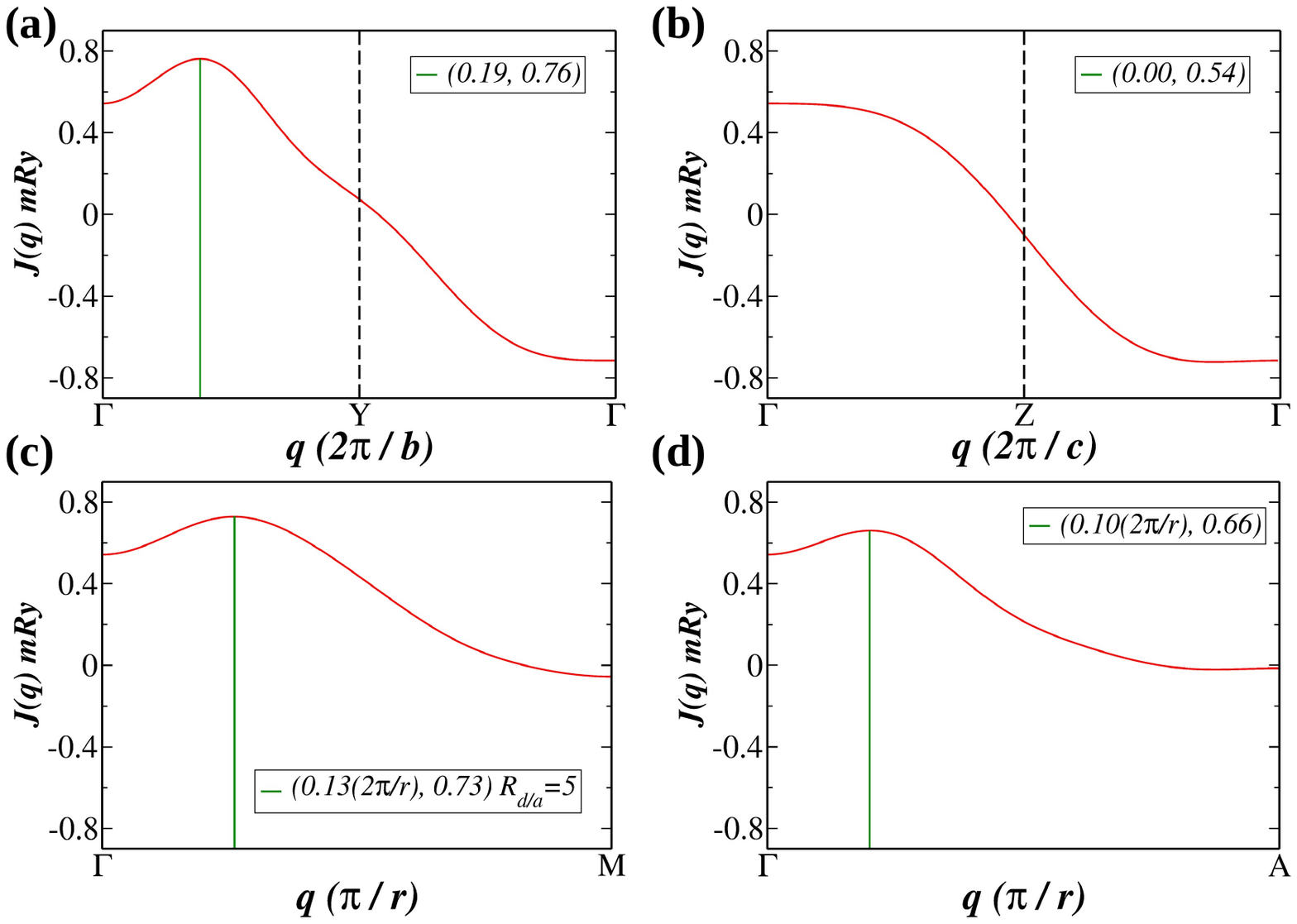}
\caption{FT $J(q)$ of the calculated exchange for $\textbf{q}$ $\parallel$ (a)$\Gamma-Y$ (b)$\Gamma-Z$ (c)$\Gamma-M$ (d)$\Gamma-A$.}
\label{ft_all_mft}
\end{figure}

The FT $J(q)$ of the calculated exchange for $\textbf{q}$ $\parallel$ different high-symmetry (HS) directions (see Fig.~\ref{bz}) of the Brillouin zone (BZ) is shown comparatively in Fig.~\ref{ft_all_mft}. The direction $\Gamma-X$ is not shown here and is shown in the main text. For this analysis, $R_{d/a} = 5$ was considered as indicated in panel (c). 

\FloatBarrier

\section{$\textbf{Q}_{111}$ modulated Spin spirals along [111] in Phase I, II, and III}

We have shown in the main text that the interlayer spin-spin correlation results in a cycloidal-type spiral along the [111] direction defined by the modulation vector $\textbf{Q}_{111}$ in Phase I. This modulation vector and the correlation also remain in Phase II and III, transforming the cycloid to a helical- and conical-type spiral, respectively, as shown in Fig.~\ref{inter-layer-spiral-all} below. Notice that the spiral rotation direction $\hat{e}_\mathrm{rot}$ for this spiral in Phase I was nearly parallel to the $[1\bar{1}0]$ direction. But in Phases II and III it is nearly $\parallel$ [111]. 

\begin{figure}[ht]
\includegraphics[scale=0.5]{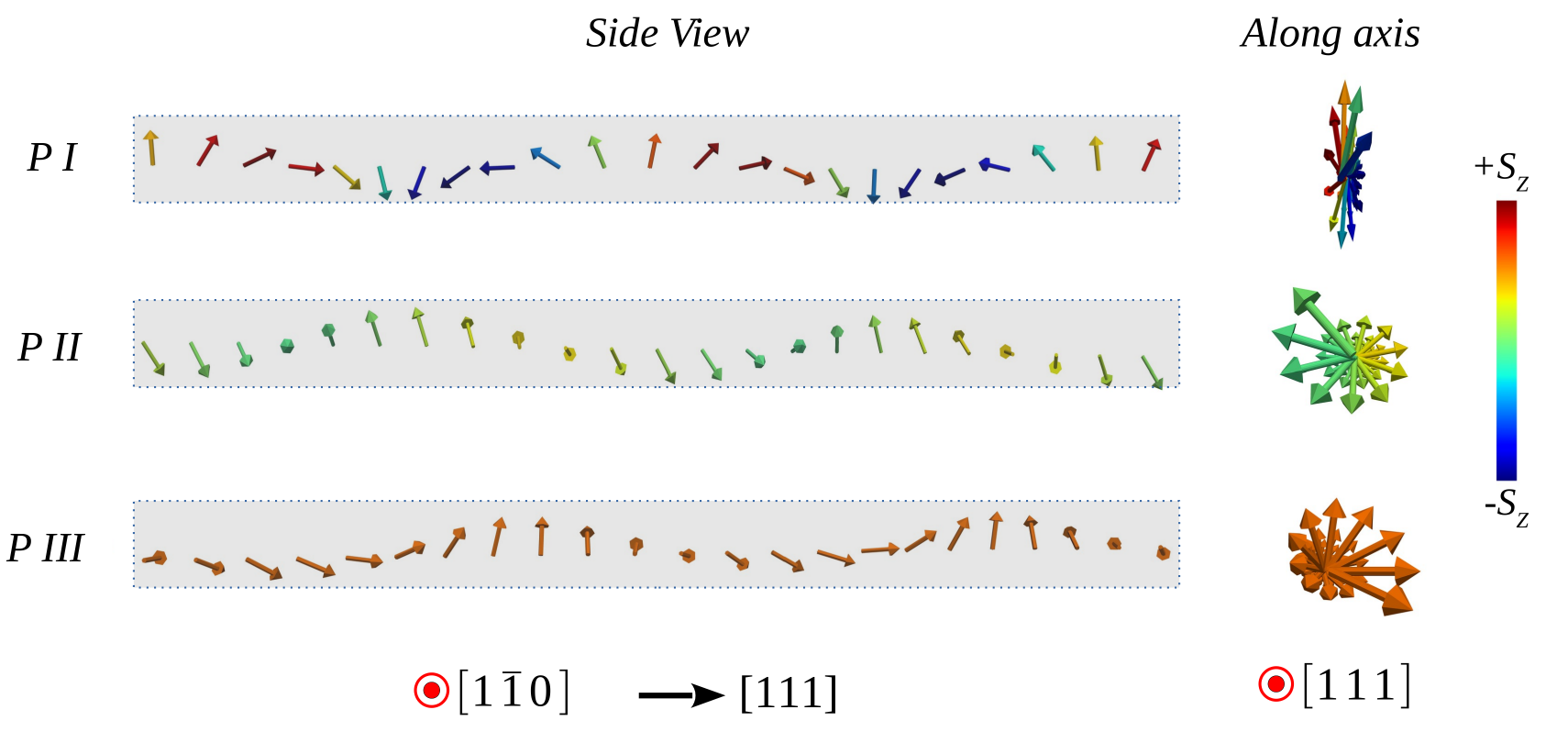}
\caption{Spin spiral on the chain of spins along the [111] direction defined by the modulation vector $\textbf{Q}_{111}$ in Phase I, II, and III. Both side-view and vertical-view along the [111] axis are shown.}
\label{inter-layer-spiral-all}
\end{figure}
\FloatBarrier

\section{The methodology for Spin Structure Factor calculation.}

The spin ordering or configurations in the 2D Gd layers in phases I, II, and III differ. The differences could be easily identified from the spin structure factor, the Fourier transform of the static spin-spin pair correlation function. This is defined as follows~\cite{hayami2021square}. 

\begin{equation}
 S^{\alpha}_s (\textbf{q}) = \frac{1}{N} \sum_{j,l}   S^{\alpha}_j S^{\alpha}_l e^{i\textbf{q} \cdot (\textbf{r}_j - \textbf{r}_l)}
 \label{eqn1}
\end{equation}

$S^{\alpha}_s$ stands for the $\alpha^\mathrm{th}$ component of the spin structure factor with $\alpha = x, y, z$. $N$ is the total number of spins in a single Gd layer. $S^{\alpha}_j$ is the $\alpha^\mathrm{th}$ component of the $j^\mathrm{th}$ spin located at site $j$ with the position vector $\textbf{r}_j$. From here we further define an in-plane (IP) and out-of-plane (OP) spin structure factor as defined in the following Eqns.~\ref{eqn2} and~\ref{eqn3}, respectively, to easily understand the spin modulations in different spin configurations. 

\begin{equation}
 S^{IP}_s (\textbf{q}) = S^{x}_s (\textbf{q}) + S^{y}_s (\textbf{q})
 \label{eqn2}
\end{equation}

\begin{equation}
 S^{OP}_s (\textbf{q}) = S^{z}_s (\textbf{q}) 
 \label{eqn3}
\end{equation}

\section{Spin Structure Factor of Phase I, Phase II, and Phase III in the $M-B$ phase diagram for $K_\mathrm{U} = \unit[0.05]{meV}$.}

\begin{figure}[ht]
\includegraphics[scale=0.75]{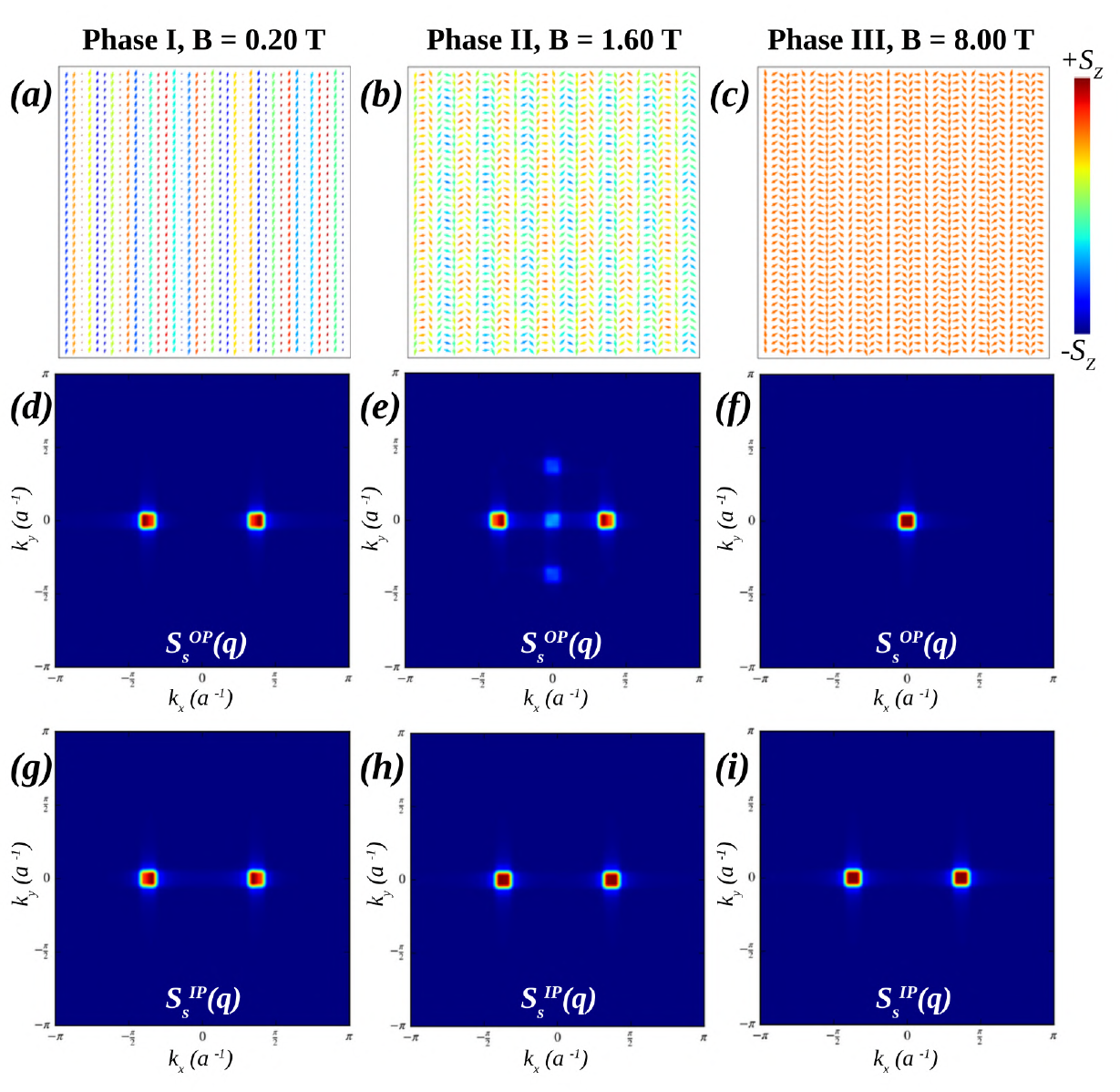}
\caption{TOP PANEL: Spin configuration within the $37\times37\times1$ 2D Gd layers in (a) phase I, (b) phase II, and (c) phase III respectively after the ASD simulation with $K_U = 0.05$ meV. MIDDLE PANEL: The out-of-plane (OP) spin structure factor ($S^{OP}_s$) defined in Eqn.~\ref{eqn2} for (d) phase I, (e) phase II, and (f) phase III, respectively. BOTTOM PANEL: The in-plane (IP) spin structure factor ($S^{IP}_s$) defined in Eqn.~\ref{eqn3} for (g) phase I, (h) phase II, and (i) phase III, respectively.}
\label{structure-factor}
\end{figure}

The calculated IP and OP Spin Structure Factor in the momentum space corresponding to phase I, phase II, and phase III in the $M-B$ phase diagram for $K_\mathrm{U} = \unit[0.05]{meV}$ is shown in Fig.~\ref{structure-factor}. 

\FloatBarrier

The IP spin structure factor of all three phases is identical showing a single-$Q$ modulation. The differences between them are, instead, visible from the OP spin structure factor which has been described in the main text in detail.

\section{$z$-component of the spin density for Phase I, Phase II, and Phase III in the $M-B$ phase diagram for $K_\mathrm{U} = \unit[0.05]{meV}$.}

\begin{figure}[ht]
\includegraphics[scale=0.55]{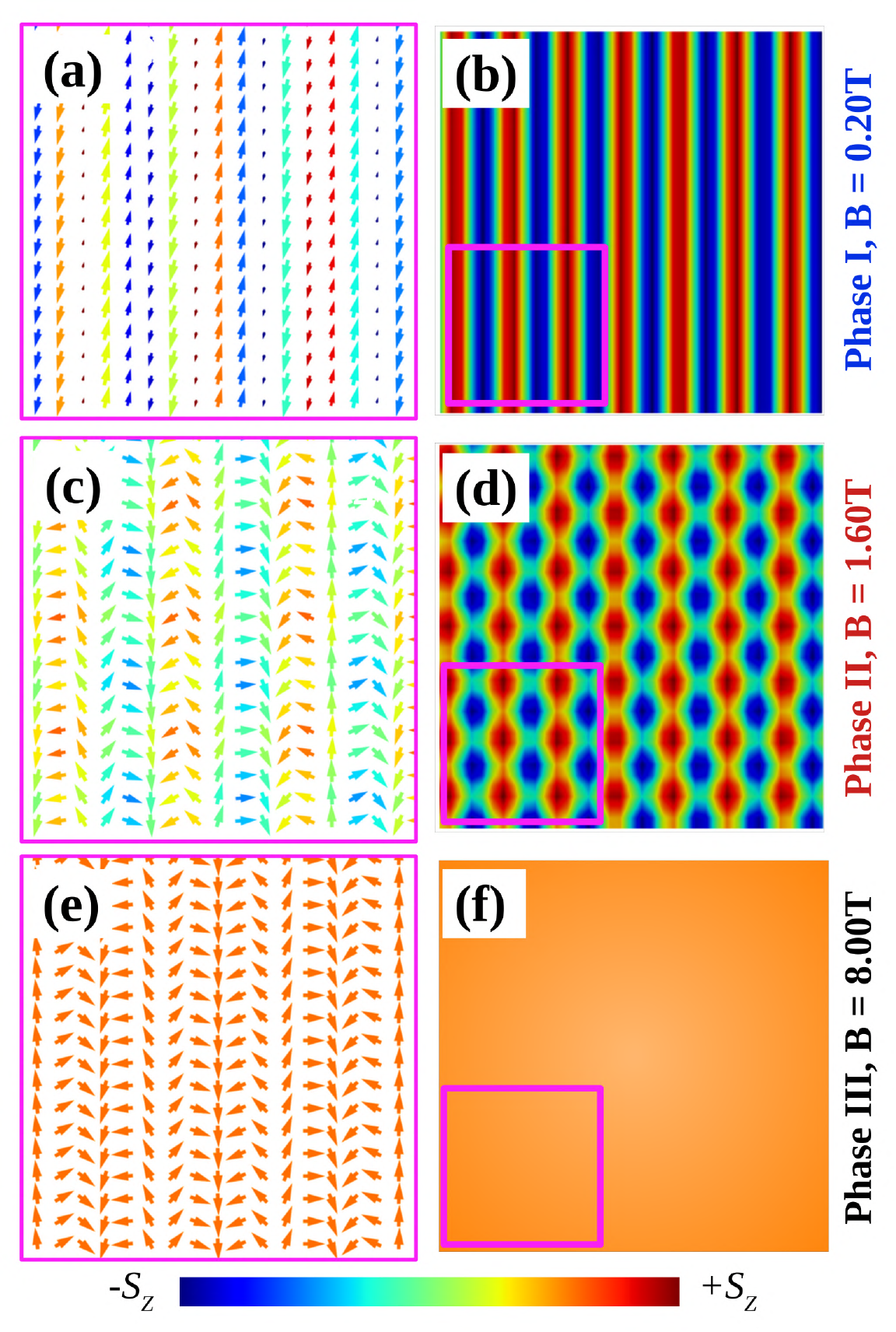}
\caption{LEFT COLUMN: Spin configuration within the 2D Gd layers in (a) Phase I, (c) Phase II, and (e) Phase III, respectively, after the ASD simulation with $K_\mathrm{U} = \unit[0.05]{meV}$. A $17\times17\times1$ square was cut from the $37\times37\times1$ plane (see the right columns) for a better view of the arrangements. RIGHT COLUMN: $S_z$ component of a spin-density function replacing the individual spins within the 2D Gd layers in (b) Phase I, (d) Phase II, and (f) Phase III respectively. In Phase II we get a square lattice distribution of $+S_z$ and $-S_z$ densities similar to the SkL state.}
\label{spindensity}
\end{figure}
\FloatBarrier

The spin configuration within a 2D Gd plane is presented in the left column of Fig.~\ref{spindensity}. In Phase I we have a helical spiral and a conical spiral in Phase III.  Phase II is neither Phase I nor Phase III but something different which becomes more evident when we try to compare the real space spin-density distribution of the $z$-component of the spin density (Fig.~\ref{spindensity} (b), (d), and (f)). In Phase II we see a unique square lattice distribution of spin-up and spin-down densities in a checkerboard manner which cannot be found in Phases I and III. Interestingly, this square-lattice distribution of $S_z$ density is also something to expect in the measured SkL phase. This happens due to a double-$\textbf{Q}$ ($\textbf{Q}_{100}$ + $\textbf{Q}_{010}$) modulation of $S_z$ that we can see from the spin structure factor in Fig.S~\ref{structure-factor} (e) of the SM, again very similar to what happens in the SkL phase. However, a similar double-$\textbf{Q}$ modulation of ($S_x + S_y$) expected for that SkL state~\cite{hayami2021square} is missing in our simulated Phase II, and this fact prevents the SkL from appearing fully.

\section{$42\times42\times42$ simulation cell calculation: Spin configurations and structure factor of Phase I, Phase II, and Phase III for $K_\mathrm{U} = \unit[0.05]{meV}$.}

\begin{figure}[ht]
\includegraphics[scale=0.75]{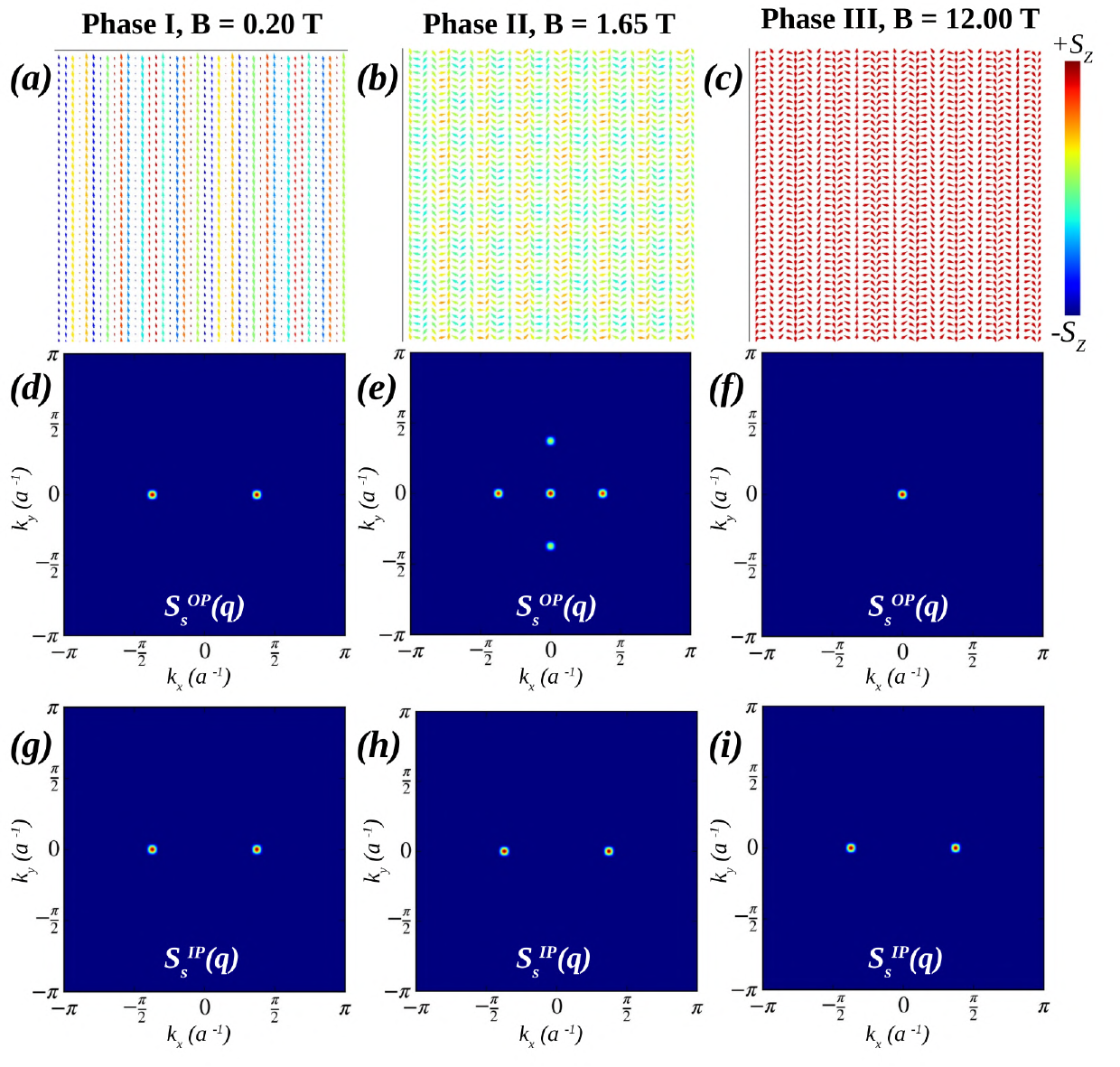}
\caption{TOP PANEL: Spin configuration within the $42\times42\times1$ 2D Gd layers in (a) phase I, (b) phase II, and (c) phase III respectively after the ASD simulation with $K_U = 0.05$ meV. MIDDLE PANEL: The out-of-plane (OP) spin structure factor ($S^{OP}_s$) defined in Eqn.~\ref{eqn2} for (d) phase I, (e) phase II, and (f) phase III, respectively. BOTTOM PANEL: The in-plane (IP) spin structure factor ($S^{IP}_s$) defined in Eqn.~\ref{eqn3} for (g) phase I, (h) phase II, and (i) phase III, respectively.}
\label{424242}
\end{figure}
\FloatBarrier

\section{Total energy method to calculate $J_1$, $J_2$, and $J_3$ in RSPt and VASP}

\begin{figure}[ht]
\includegraphics[scale=0.60]{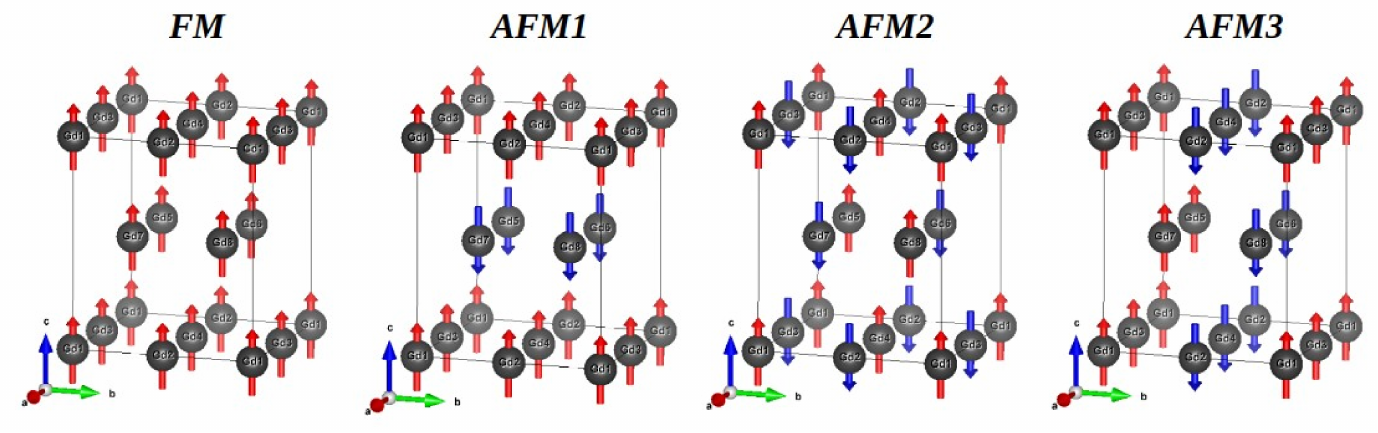}
\caption{FM and three different AFM ordering of the Gd spins in a $2\times2\times1$ supercell of tetragonal GdRu$_2$Si$_2$.}
\label{vasp-spinorder}
\end{figure}

For this, ferromagnetic (FM) and three antiferromagnetic (AFM) configurations were considered in a $2\times2\times1$ supercell of tetragonal GdRu$_2$Si$_2$ as shown in Fig.~S\ref{vasp-spinorder}.

\FloatBarrier

Total energies were calculated for these magnetic supercells and then mapped on the following Heisenberg Hamiltonian with only the first, second and third neighbor interactions $J_1$, $J_2$, $J_3$, 

\begin{equation}
 {H} = - 1/2 \sum_{i\neq j} J_{ij}\, \vec{e}_{i}\cdot\vec{e}_{j}.
 \label{eqn4}
\end{equation}

Here, $(i,j)$ are the indices for the magnetic sites and $\vec{e}_i$ and $\vec{e}_j$ are the unit vectors along the spin directions at sites $i$, and $j$, respectively. $(i,j)$ runs from 1 to 3. Total energy mapping leads to three algebraic equations with three unknown parameters. These were solved to determine the values of the parameters $J_1$, $J_2$, and $J_3$. Total energies were calculated in RSPt and VASP, and the calculated values of the parameters are reported in Table~II of the main text. The total energy calculated values of $J_1$, $J_2$, and $J_3$ from RSPt and VASP are very close to each other. Hence, for any further calculations, the VASP calculated values were considered. 

\begin{figure}[ht]
\includegraphics[scale=0.50]{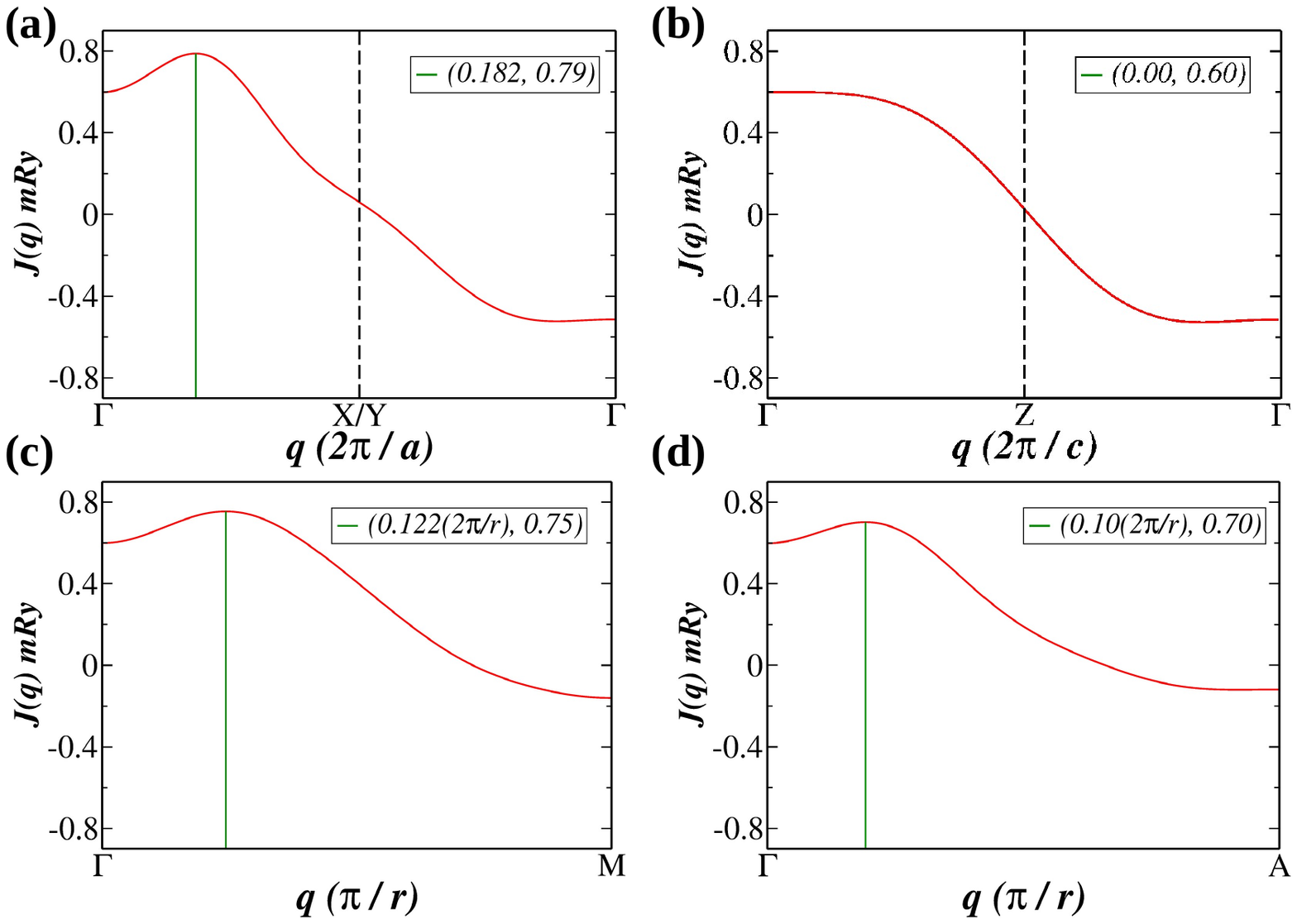}
\caption{FT $J(q)$ of $J_{ij}$ with VASP calculated $J_1$, $J_2$, and $J_3$ for $\textbf{q}$ $\parallel$ (a)$\Gamma-X/Y$ (b)$\Gamma-Z$ (c)$\Gamma-M$ (d)$\Gamma-A$. Due the square lattice symmetry of the 2D Gd plane, $J(q)$ along $\Gamma-X$ and $\Gamma-Y$ are identical.}
\label{vasp-ftall}
\end{figure}

To check the role of the new exchange values, FT $J(q)$ of the exchange parameters $J_{ij}$ was calculated with $J_1$, $J_2$ and $J_3$ from the VASP total energy calculation. $J(q)$ along different direction of the Brillouin zone is shown in Fig.~S\ref{vasp-ftall}. 

\FloatBarrier

\begin{figure}[ht]
\includegraphics[scale=0.70]{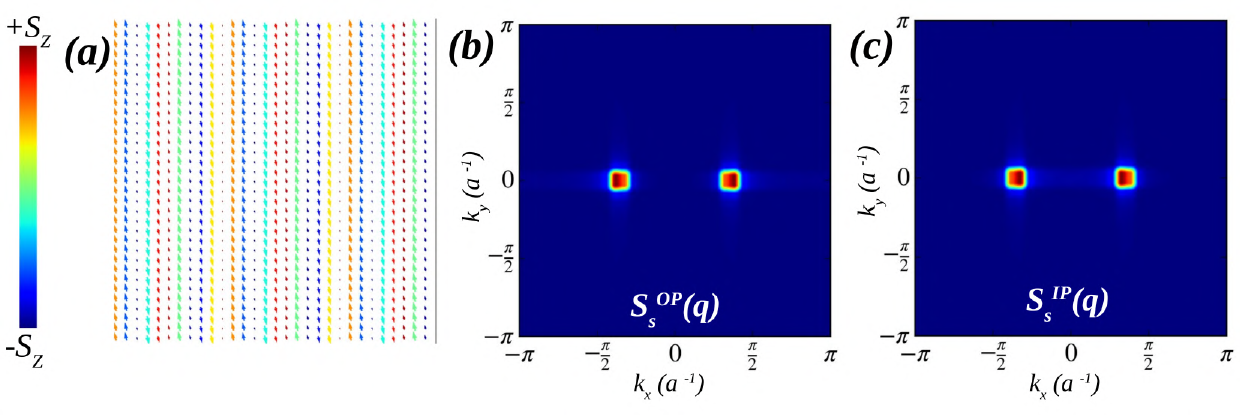}
\caption{(a) Spin configuration within the 2D Gd layers in the zero-field ground state. (b) The OP spin structure factor showing a single-$\textbf{Q}$ modulation. (c) The IP spin structure factor showing a single-$\textbf{Q}$ modulation.}
\label{vasp-spinconfig}
\end{figure}

After this, ASD simulations were done considering the above exchange parameters along with $K_U$ = 0.05 meV to determine the zero-field magnetic ground state. A helical spiral state was stabilized with a single-$\textbf{Q}$ modulation. The spin configuration in the 2D Gd layers and IP, OP spin structure factors are shown in Fig.~S\ref{vasp-spinconfig}. Based on the magnitude of $\textbf{Q}_{100}$, a $33\times33\times33$ simulation cell was used for this calculation.

\FloatBarrier

\section{Spin configurations and Structure Factor of Phase I, Phase II, and Phase III for $K_\mathrm{U} = \unit[0.10]{meV}$.}

\begin{figure}[ht]
\includegraphics[scale=0.75]{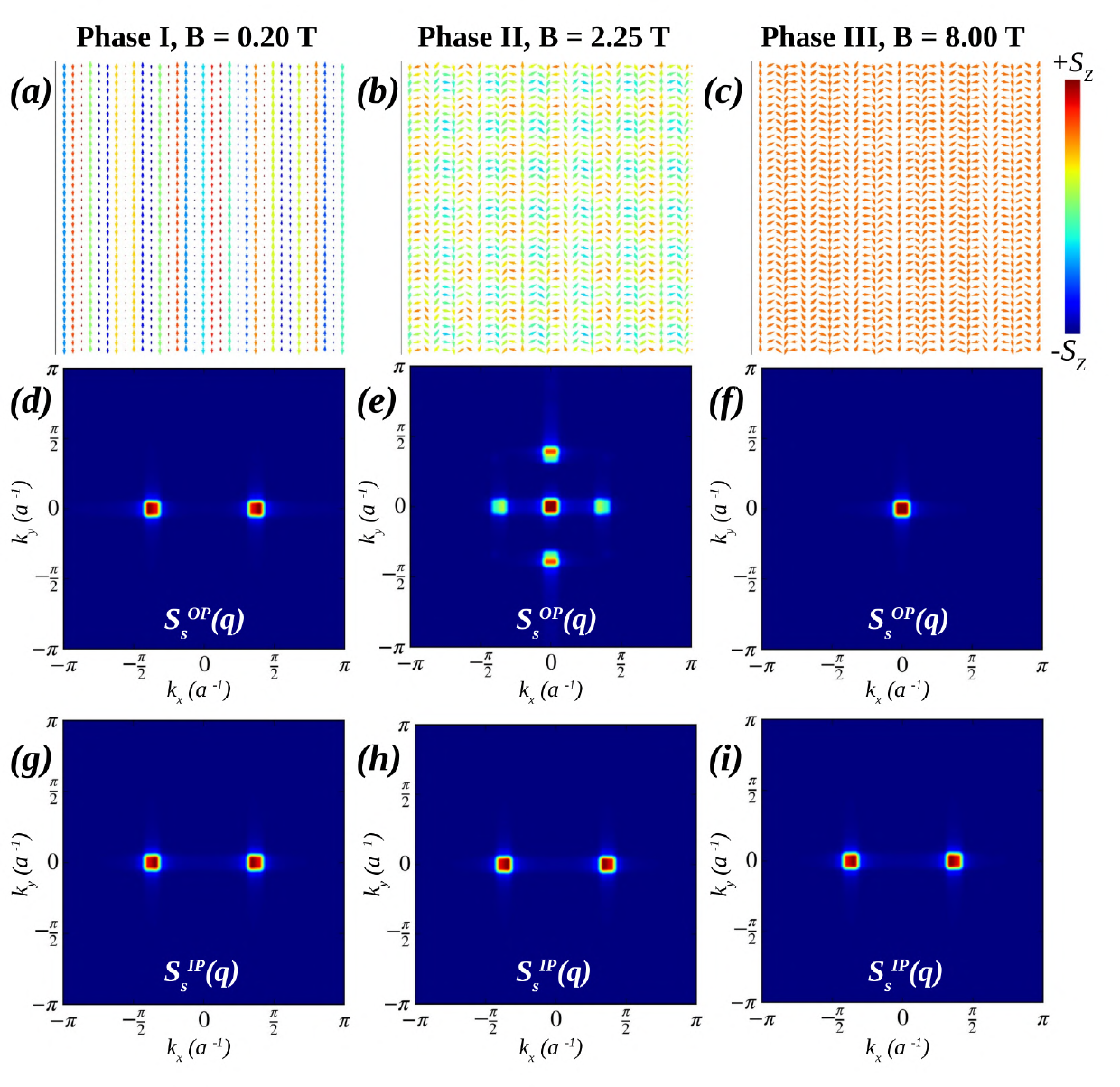}
\caption{TOP PANEL: Spin configuration within the $37\times37\times1$ 2D Gd layers in (a) phase I, (b) phase II, and (c) phase III respectively after the ASD simulation with $K_\mathrm{U} = \unit[0.10]{meV}$. MIDDLE PANEL: The OP spin structure factor ($S^{OP}_s$) defined in Eqn.~\ref{eqn2} for (d) phase I, (e) phase II, and (f) phase III respectively.  BOTTOM PANEL: The IP spin structure factor ($S^{IP}_s$) defined in Eqn.~\ref{eqn3} for (g) phase I, (h) phase II, and (i) phase III, respectively.}
\label{k0p10_phases}
\end{figure}

The spin configuration within a 2D Gd layer corresponding to phase I, phase II, and phase III is shown in Fig.~\ref{k0p10_phases} (a-c). The corresponding IP and OP spin structure factors for the three phases are also presented in the middle and bottom panels of Fig.~\ref{k0p10_phases} respectively.

\FloatBarrier

\section{Effect of uniaxial anisotropy $K_\mathrm{U}$ in the presence of dipolar interactions.}

\begin{figure}[ht]
\includegraphics[scale=0.55]{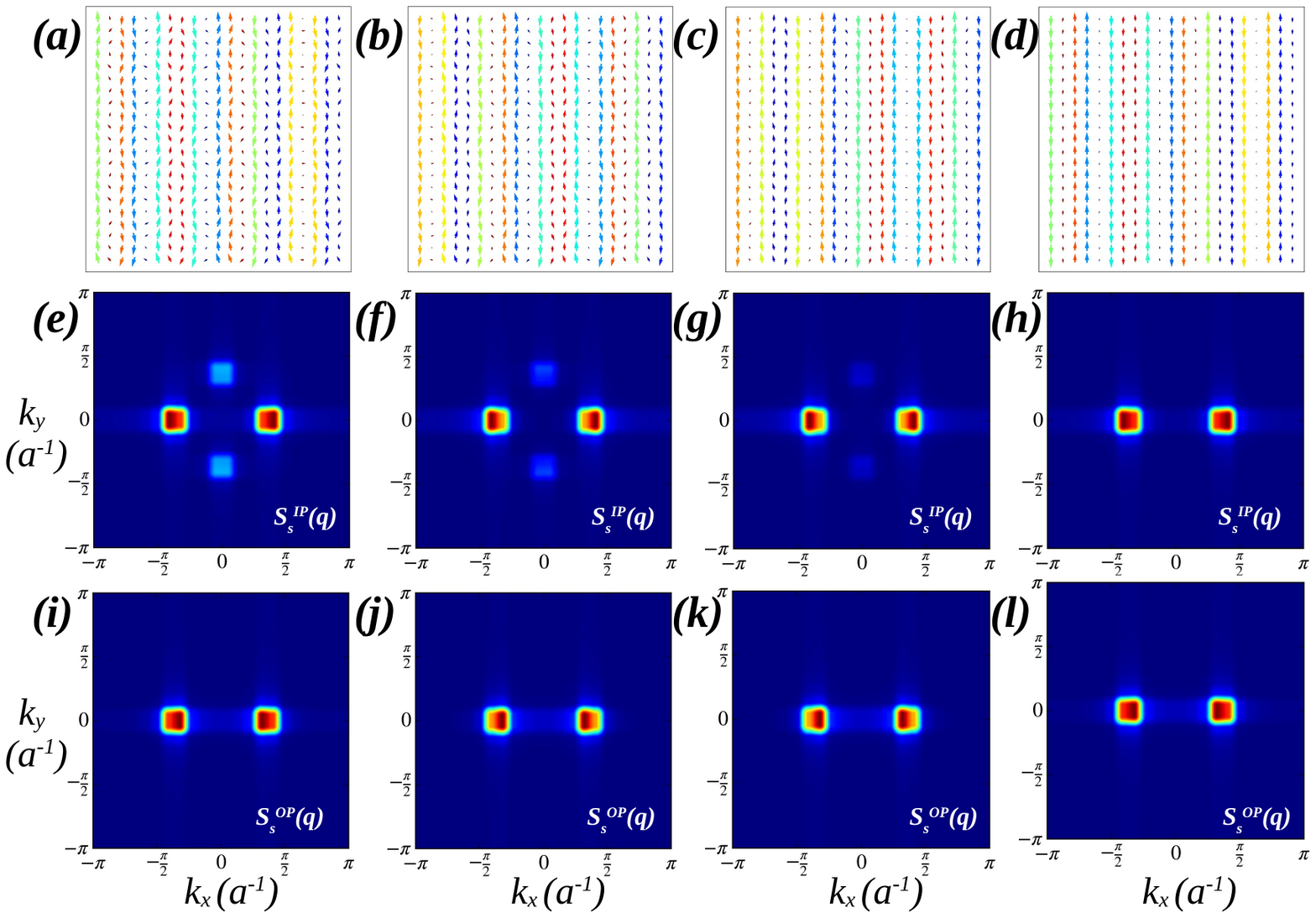}
\caption{Effect of uniaxial anisotropy $K_\mathrm{U}$ in the presence of dipolar interactions. TOP PANEL: Spin configuration within the $21\times21\times1$ 2D Gd layers for (a) $K_\mathrm{U} = 0$, (b) $K_\mathrm{U} = 0.05$,  (c) $K_\mathrm{U} = 0.10$ and (d) $K_\mathrm{U} = 0.15$ meV respectively. MIDDLE PANEL: The corresponding IP spin structure factor ($S^{IP}_s$)  of the spin states shown in the top panel. BOTTOM PANEL: The corresponding OP spin structure factor ($S^{OP}_s$)  of the spin states shown in the top panel.}
\label{dip-role-full}
\end{figure}

Spin configurations and spin structure factor of zero-field ground states in the presence of dipolar interactions and for different values of the uniaxial anisotropy $K_\mathrm{U}$ are shown in Fig.~S\ref{dip-role-full}. See the main text for a description of the data.

\FloatBarrier

\section{ASD simulations in the presence of dipolar interactions and without $K_\mathrm{U}$}. 

\begin{figure}[ht]
\includegraphics[scale=0.75]{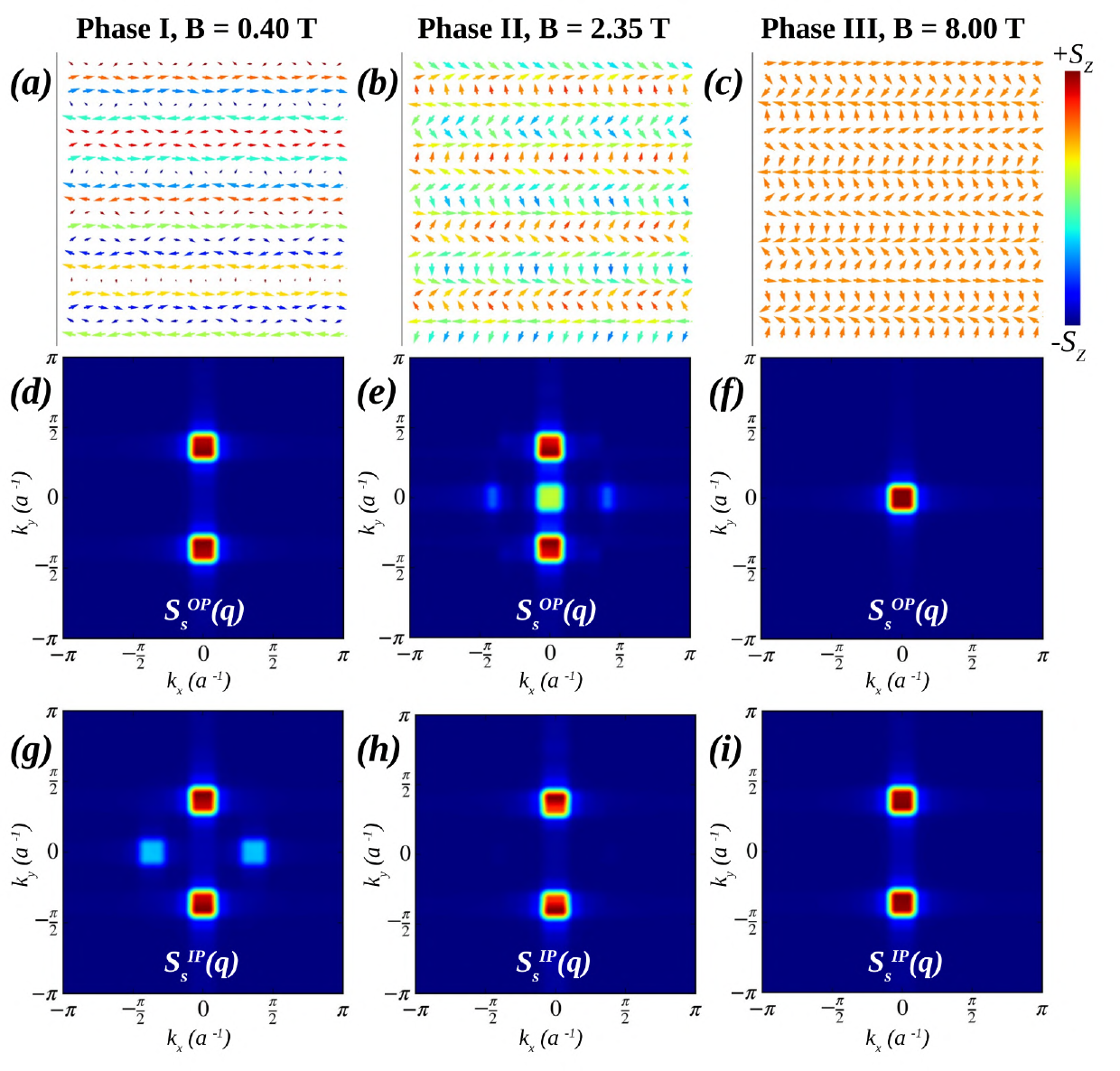}
\caption{TOP PANEL: Spin configuration within the $21\times21\times1$ 2D Gd layers in (a) phase I, (b) phase II, and (c) phase III respectively after the ASD simulation without $K_\mathrm{U}$ and in the presence of dipolar interactions. MIDDLE PANEL: The OP spin structure factor ($S^{OP}_s$) defined in Eqn.~\ref{eqn2} for (d) phase I, (e) phase II, and (f) phase III respectively.  BOTTOM PANEL: The IP spin structure factor ($S^{IP}_s$) defined in Eqn.~\ref{eqn3} for (g) phase I, (h) phase II, and (i) phase III, respectively.}
\label{dip_phases}
\end{figure}

Spin configurations and structure factors of Phase I, Phase II, and Phase III are shown in Fig.~S\ref{dip_phases}. See the main text for details. 

\FloatBarrier

\medskip
\bibliographystyle{naturemag.bst}
\bibliography{sm}